\documentclass[useAMS,usegraphicx,usenatbib]{mn2e}
\usepackage{times}
\usepackage{rotate}
\usepackage{amssymb}
\usepackage{multirow}
\usepackage{lscape}
\usepackage{rotating}
\title[Are reddened 2MASS QSOs young?]
{Are red 2MASS QSOs young?}
\author[Georgakakis et al. ] {A. Georgakakis$^{1}$\thanks{email: age@astro.noa.gr}, D. L. Clements$^{2}$, 
  G. Bendo$^{2}$, M. Rowan-Robinson$^{2}$, K. Nandra$^{2}$, \\\\{\rm
    \LARGE M. S. Brotherton$^{3}$}\\\\
  $^1$National Observatory of Athens, I. Metaxa \& V. Paulou,
  Athens 15236, Greece\\ 
  $^2$Astrophysics Group, Blackett Laboratory, Imperial College, Prince
  Consort Rd , London SW7 2BZ, UK\\
  $^3$Department of Physics and Astronomy, The University of Wyoming
  (Department 3905), 1000 East University Avenue, Laramie, WY 82071 
}
\begin{document}
\maketitle  

\begin{abstract} 
We  use photometric data from the  Spitzer mission to
explore the mid- and far-infrared  properties of 10 red QSOs ($J-K>2$,
$R-K>5$\,mag)  selected by  combining the  2MASS in  the near-infrared
with  the SDSS  at optical  wavelengths. Optical  and/or near-infrared
spectra are available for  8/10 sources.  Modeling the Spectral Energy
Distribution (SED)  from UV to  far-infrared shows that  moderate dust
reddening   ($A_V = 1.3 - 3.2$) can explain  the red  optical  and near-IR  colours of  the
sources in the sample. There is  also evidence that red QSOs have $\rm
60/12\mu  m$  luminosity  ratio  higher  than PG  QSOs  (97  per  cent
significance).   This  can  be  interpreted as  a  higher  level  of
star-formation  in these  systems (measured  by the  $\rm 60  \,\mu m$
luminosity) for a  given AGN power (approximated by  the $\rm 12 \,\mu
m$  luminosity).  This  is consistent  with a  picture where  red QSOs
represent an early phase of AGN evolution, when the supermassive black
hole is  enshrouded in dust and  gas clouds, which  will eventually be
blown out (possibly by AGN driven outflows) and the system will appear
as typical  optically luminous QSO.  There is  also tentative evidence
significant  at the  96\% level  that red  2MASS QSOs  are  more often
associated  with radio  emission  than optically  selected SDSS  QSOs.
This  may  indicate  outflows,  also  consistent with  the  young  AGN
interpretation.   We  also estimate  the  space  density  of red  QSOs
relative  to optically  selected SDSS  QSOs, taking  into  account the
effect of dust extinction and the intrinsic luminosity of the sources.
We estimate  that the fraction of  red QSOs in  the overall population
increases from  3\% at  $M_K=-27.5$\,mag to 12\%  at $M_K=-29.5$\,mag.
This suggests that  reddened QSOs become more important  at the bright
end of the Luminosity Function.  If red QSOs are transition objects on
the way to becoming typical optically luminous QSOs, the low fractions
above  suggest  that these  systems  spent  less  than 12\%  of  their
lifetime at the ``reddened'' stage.
\end{abstract}
\begin{keywords}  
  Surveys  --  galaxies:  active  --  galaxies: quasars: general
  -- infrared: general
\end{keywords}

\section{Introduction}\label{sec_intro}  In  the  last  few  years  an
increasing body  of evidence  points to an  intimate link  between the
formation of galaxies  and the growth of the  supermassive black holes
(SBH) at  their centres.   For example, the  accretion history  of the
Universe (Barger et al.  2005; Hasinger et al.  2005) looks strikingly
similar   to  the   evolution  of   the  cosmic   star-formation  rate
\citep[e.g.][]{Hopkins_A2004,   Hopkins_A2006}.    Additionally,  most
nearby  spheroids contain  a  SBH \citep[SBH;  e.g.][]{Magorrian1998},
with a  mass that scales with  the stellar velocity  dispersion of the
host   galaxy    bulge   \citep[e.g][]{Ferrarese2000,   Gebhardt2000},
suggesting similar formation epochs and/or mechanisms.

Recent cosmological simulations account  for the observations above by
assuming  that  both  the  galaxy  and  the  SBH  form  during  galaxy
interactions  and mergers  \citep[e.g][]{Kauffmann2000, Kauffmann2002,
DiMatteo2003, Menci2004, Cattaneo2006}.   Such violent events are very
efficient in driving gas to  the nuclear regions, where it forms stars
or it  is consumed  by the  SBH.  Because the  nuclear regions  of the
interacting system  are dusty, both  processes above most  likely take
place within a cocoon of dust and gas clouds. Energetic considerations
underline the  significance of the energy  released by the  AGN in the
process    above,    which    can    regulate    the    gas    inflows
\citep[e.g][]{Silk1998,  Fabian1999, King2003}.  Simulations  of major
mergers indeed show that the feedback from the AGN can be particularly
violent close  to the final  merging, leading to  the blow out  of the
dust and gas clouds and the termination of both the star-formation and
eventually the AGN activity \citep[e.g.][]{DiMatteo2005, Springel2005,
Hopkins2006}.

The key  prediction of the  picture above is  that there should  be an
association  between SBH accretion,  starbursts and  mergers. Although
there   are  examples   of  such   systems  \citep[e.g.][]{Genzel1998,
  CidFernandes2001, Franceschini2003, Georgakakis2004, Alexander2005},
they appear to be the  exception rather than the rule.  X-ray surveys,
arguably the  most efficient method  for finding AGN, have  shown that
the  majority of  X-ray sources  are  hosted by  red evolved  galaxies
\citep{Nandra2007},  with prominent  bulges  and little  morphological
evidence  for  ongoing  interactions  \citep{Grogin2003,  Pierce2007}.
This  may  suggest  that  X-ray  wavelengths, because  of  their  high
sensitivity to AGN, select a large number of systems after the peak of
their activity (i.e.   final merging) and during the  decline stage of
the SBH accretion.   In order to understand the  interplay between SBH
and  galaxy formation  one should  study the  properties of  AGN hosts
(e.g. star-formation  rate, optical morphology)  close to the  peak of
the activity, when  the system appears as luminous  QSO.  Such studies
however, are  hampered by  the enormous brightness  of the  QSO nuclei
relative to the  host galaxy.  Nevertheless progress has  been made in
the last  few years.   Mid-Infrared spectroscopy with  Spitzer reveals
PAH  features, i.e.   evidence for  star-formation, in  about  30\% of
optically  selected luminous QSOs  \citep{Schweitzer2006, Netzer2007}.
HST  high  resolution  imaging  and  ground  based  observations  with
adaptive optics find  that a fraction of the  QSO population is hosted
by either  elliptical galaxies with fine structure  indicative of past
interactions  \citep[80\%;][]{Bennert2008}, or  galaxies  with obvious
signs of  disturbance \citep[30\%;][]{Guyon2006}.  The  evidence above
suggests  that  at  least  some  QSOs  are  formed  in  major  mergers
accompanied by starburst events.
 
An important development  in studies of the interplay  between AGN and
galaxies  has been  the  identification of  QSOs  with optical  and/or
near-infrared  (near-IR) colours  much  redder than  those of  typical
UV/optically  selected QSOs.   \cite{Cutri2001}  and \cite{Wilkes2002}
for example,  have found a  previously undetected population  of broad
line AGN  among sources  with $J-K>2$\,mag in  the Two Micron  All Sky
Survey  (2MASS).   \cite{White2003}   studied  an  $I$-band  magnitude
limited ($I<20.5$\,mag)  sample of radio  sources and showed  that the
QSOs in that sample have  redder $B-R$ colours, on average, than those
selected  at bluer  optical filters.   \cite{Glikman2004, Glikman2007}
demonstrated that selecting radio  (1.4\,GHz) sources in the region of
the colour  space $J-K>1.7$  and $R-K>4.0$\,mag yields  a 50  per cent
success rate  of discovering QSOs  that are substantially  redder than
those  found in optical  surveys. Similar  results were  obtained more
recently by \cite{Urrutia2008a} using a slightly modified colour wedge
for  selecting  radio  sources  ($J-K>1.3$ and  $r-K>5.0$\,mag).   The
optical/near-IR continua  of the red QSO population  identified in the
studies above  are consistent with moderate amounts  of dust reddening
\citep[$A_V\approx1.5-3$][]{White2003, Glikman2007, Urrutia2008a}.  An
intrinsically  flat UV/optical spectrum  \citep[e.g.][] {Richards2003}
or   synchrotron    emission   with   a    high   turnover   frequency
\citep{Whiting2001} can also explain  the red colours without invoking
dust.  It is unlikely  however, that these alternative scenarios apply
to the entire  population of red QSOs.  It  is therefore proposed that
this new  population of dust  reddened QSOs represent  systems shortly
before or during the blow out stage of AGN evolution, when the central
source is  still cocooned in dust  and gas clouds.   If true, reddened
QSOs  are  intermediate  between  dusty  starbursts  and  UV/optically
luminous QSOs. Clues on the evolutionary stage of these systems can be
obtained  by studying  the properties  of their  host  galaxies.  Deep
optical imaging for example, has  shown that the morphology of the red
QSO hosts are in most cases disturbed, suggesting a higher fraction of
ongoing  interactions compared to  typical UV/optically  luminous QSOs
\citep{Hutchings2003, Hutchings2006, Urrutia2008}, consistent with the
young AGN picture.

Another property  of red QSOs that  has not yet been  addressed and is
particularly  relevant to  their evolutionary  stage is  the  level of
star-formation rate in the host galaxy.  In this paper we address this
issue using Spitzer to constrain and to model the mid- and far-IR SEDs
of a  sample of red QSOs identified  in the Two Micron  All Sky Survey
(2MASS).  Under reasonable  assumptions, the shape of the  IR SED is a
powerful diagnostic  of the physical processes  (star-formation vs AGN
activity) responsible  for heating  the dust in  extragalactic sources
\citep[e.g.][]{rowan2008}.

The  fraction  of  red QSOs  in  the  overall  population is  also  an
important parameter.  In the  scenario where these systems represent a
stage  of the  AGN phenomenon,  their fraction  provides clues  on the
timescale of this stage.  Currently, there is large uncertainty in the
fraction  of red QSOs  missing from  optically selected  samples, with
values ranging from 15  to over 50 per cent \citep[e.g.][]{Wilkes2002,
  Richards2003, White2003,  Glikman2004, Glikman2007}. The uncertainty
is associated with the  selection of the appropriate comparison sample
of  optically selected QSOs  and with  difficulties in  accounting for
dust extinction  in the selection of  reddened QSOs. In  this paper we
address these  issues by  using the 1/Vmax  formalism to  estimate the
maximum volume  that a source  is detectable, given the  survey limits
and  taking  into  account  the  effect of  dust  extinction  and  the
intrinsic luminosity  of the source.  For comparison we use  the Sloan
Digital   Sky  Survey   (SDSS)  optically   selected  QSO   sample  of
\cite{Schneider2005} , which is  large and with well defined selection
criteria.  Throughout  the paper  we adopt  $\rm H_{0} =  70 \,  km \,
s^{-1} \, Mpc^{-1}$, $\rm  \Omega_{M} = 0.3$ and $\rm \Omega_{\Lambda}
= 0.7$.

\section{Sample selection}\label{sec_sample}

Candidates for  dust enshrouded active  galaxies were selected  in the
All  Sky Data  Release  of the  2MASS  Point Source  Catalog (PSC)  by
cross-correlating the positions of the 2MASS sources with the 3rd data
release (DR3) of the SDSS.

First,  2MASS-PSC sources  with $Ks<14.5$\,mag  and $J-Ks  > 1.5$\,mag
were  selected.  The  magnitude  limit is  close  to the  99 per  cent
completeness          limit          of         the          2MASS-PSC
\citep[$Ks=14.3$\,mag;][]{Cutri2005}, while the colour cut is to avoid
Galactic  stars \citep[e.g.][]{Francis2004}.   We exclude  sources for
which   the  $Ks$-band  photometric   quality  flag   ({\sc  PH\_QUAL}
parameter) has  values ``X'' and  ``U'', which signify  bad photometry
and   an  upper   limit  in   the  $Ks$-band   magnitude  respectively
\citep{Cutri2005}.

\begin{figure}
\begin{center}
 \rotatebox{0}{\includegraphics[height=0.9\columnwidth]{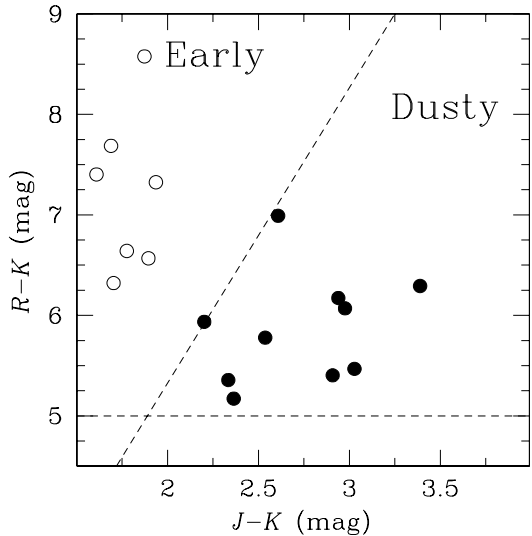}}
\end{center}
\caption{$J-K$  against $R-K$  colour plot  for the  2MASS/SDSS sample
  with $K<14.5$\,mag . The sample of dusty active galaxy candidates is
  shown with filled circles. The dashed lines show the colour criteria
  used  to select  these sources.  The horizontal  line  corresponds to
  $R-K=5$\,mag and  the diagonal line has been  introduced by Pozzetti
  \& Mannucci  (2000) to separate dusty from  early-type Extremely Red
  Objects.}\label{fig_rjk} 
\end{figure}

The resulting 2MASS-PSC source list was cross-correlated with the SDSS
using a  matching radius of 3\,arcsec,  which is much  larger than the
sub-arcsec positional accuracy of the SDSS and the 2MASS surveys.  For 
the SDSS  magnitude limit of  $r\approx23$\,mag we expect about  2 per
cent  spurious alignment  rate  within this  search  radius. The  SDSS
$r$-band  AB magnitudes  were  converted to  Vega $R$-band  magnitudes
using the colour transformations of  Fukugita et al.  (1996).  This is
to  select candidates for  dust enshrouded  active systems  using the
colour cut $(R-Ks)_{Vega}>5$\,mag, i.e.  similar to that of Extremely
Red Objects (EROs; Elston,  Rieke \& Rieke 1988).  \cite{Pozzetti2000}
have used  the $R-K$ vs  $J-K$ colour diagram to  discriminate between
early-type and dusty EROs.  This is shown in Figure \ref{fig_rjk}.  We
focus on the dusty subregion of  this figure. A total of 10 sources of
our SDSS/2MASS sample lie in the dusty part of the colour-colour space
of Figure \ref{fig_rjk}.  These  are presented in Table \ref{tab_obs1}
and form the main sample of  this paper. The offsets between the 2MASS
and the SDSS  source positions are $<$0.4\,arcsec for  all 10 selected
sources and  therefore the probability of  spurious identifications is
small,   $<0.04$  per   cent,  for   the  SDSS   magnitude   limit  of
$r\approx23$\,mag. We note that  the colour selection criteria adopted
here    are   similar   to    those   used    by   \cite[][$B-Ks>4.3$,
  $J-K>2$\,mag]{Wilkes2002}           and          \cite[][$R-Ks>4.0$,
  $J-K>1.7$\,mag]{Glikman2004} to select 2MASS red AGN.

\begin{table*} 
\caption{The 2MASS selected active dusty galaxy candidates. The columns are: 
(1): source name; 
(2): right ascension of the optical source; 
(3): declination  of the optical source; 
(4): $u$-band AB magnitude and error;
(5):  $g$-band AB magnitude and error;  
(6): $r$-band AB magnitude and error; 
(7):  $i$-band AB magnitude and error;
(8):  $z$-band AB magnitude and error;
(9):  $J$-band Vega magnitude and error;
(10): $H$-band Vega magnitude and error;
(11): $Ks$-band Vega magnitude and error;
}
\centering
\scriptsize
\begin{tabular}{l cc ccccc ccc}
\hline 
ID &
$\alpha$ & 
$\delta$ & 
$u$    &
$g$    & 
$r$    &
$i$    & 
$z$    &

$J$    &
$H$    & 
$K$    \\

\hline
2MASS\_02 & 20h56m29.76s &  -06d50m55.4s & 
$21.75\pm0.36$ & $20.34\pm0.04$ & $19.59\pm0.02$ & $18.69\pm0.01$ & $18.47\pm0.04$ & $<16.48$ & $15.54\pm0.09$ & $13.76\pm0.04$ 
\\

2MASS\_03 & 20h48m37.25s &  -00d24m37.3s  &  
$20.48\pm0.13$ & $20.12\pm0.03$ & $19.41\pm0.02$ & $18.93\pm0.02$ & $18.32\pm0.04$ & $16.67\pm0.15$ & $15.08\pm0.08$ & $13.64\pm0.04$
\\

2MASS\_04 & 17h04m21.16s & +24d33m40.9s &
$22.40\pm0.42$ & $21.57\pm0.07$ & $20.23\pm0.03$ & $19.36\pm0.02$ & $18.65\pm0.05$ & $16.82\pm0.17$ & $15.72\pm0.13$ & $14.49\pm0.08$
\\

2MASS\_05 & 15h33m34.18s & +07d35m05.7s  &
 $22.12\pm0.94$ & $22.66\pm0.50$ & $20.97\pm0.16$ & $19.65\pm0.07$ & $18.60\pm0.11$ & $16.98\pm0.23$ & $15.82\pm0.14$ & $14.44\pm0.11$
\\

2MASS\_06 &  13h40m39.68s & +05d14m19.9s &
 $20.69\pm0.10$ & $19.82\pm0.02$ & $18.97\pm0.01$ & $18.46\pm0.01$ & $17.70\pm0.02$ & $16.16\pm0.11$ & $14.80\pm0.08$ & $13.25\pm0.03$
\\

2MASS\_07 & 13h21m22.89s & +45d02m24.8s &
$24.52\pm2.26$ & $22.44\pm0.22$ & $20.75\pm0.06$ & $19.36\pm0.03$ & $18.61\pm0.05$ & $16.58\pm0.13$ & $15.12\pm0.08$ & $14.38\pm0.09$
\\

2MASS\_08 &  12h52m12.93s&  +07d15m04.7s &
 $21.92\pm0.33$ & $21.30\pm0.06$ & $21.03\pm0.07$ & $20.30\pm0.05$ & $19.43\pm0.09$ & $<17.88$ & $16.53\pm0.18$ & $14.49\pm0.10$
\\

2MASS\_10 & 08h25m2.06s &  +47d16m52.0s & 
$21.38\pm0.29$ & $20.88\pm0.07$ & $20.50\pm0.07$ & $19.68\pm0.05$ & $19.56\pm0.16$ & $17.14\pm0.22$ & $15.84\pm0.18$ & $14.16\pm0.06$
\\

2MASS\_11 & 01h34m35.67s &  -09d31m03.0s &
 $25.52\pm0.81$ & $23.51\pm0.32$ & $21.20\pm0.05$ & $19.60\pm0.02$ & $18.31\pm0.03$ & $16.18\pm0.13$ & $14.79\pm0.07$ & $13.58\pm0.05$
\\

2MASS\_12 & 00h36m59.82s & -01d13m32.4s &
 $22.05\pm0.31$ & $21.52\pm0.07$ & $20.35\pm0.03$ & $19.88\pm0.03$ & $18.66\pm0.04$ & $16.56\pm0.10$ & $15.10\pm0.06$ & $13.63\pm0.04$
\\
\hline
\end{tabular}\label{tab_obs1} 
\end{table*}

\begin{table*} 
\caption{The 2MASS selected active dusty galaxy candidates. The columns are: 
1: source name; 
2: log of IRAC $\rm 3.6\mu m $ flux density in $\rm \mu Jy$. The
error includes both background and callibration uncertainties;  
3: log of IRAC $\rm 4.5\mu m $ flux density in $\rm \mu Jy$;
4: log of IRAC $\rm 5.8\mu m $ flux density in $\rm \mu Jy$;
5: log of IRAC $\rm 8.0\mu m $ flux density in $\rm \mu Jy$;
6: log of MIPS $\rm 24\mu m $ flux density in $\rm \mu Jy$;
7: log of MIPS $\rm 70\mu m $ flux density in $\rm \mu Jy$;
8: log of MIPS $\rm 160\mu m $ flux density in $\rm \mu Jy$;
9: 1.4GHz radio flux density in mJy. All measurement are from FIRST except for sources  2MASS\_2 and 2MASS\_3, which are from the NVSS.
10: Redshift. Photometric redshifts are marked with ``ph''. See text for details.} 
\centering
\scriptsize
\begin{tabular}{l cccc ccc cc}
\hline 
ID &

$\rm 3.6\,\mu m $    & 
$\rm 4.5\,\mu m $    &
$\rm 5.8\,\mu m $    & 
$\rm 8.0\,\mu m $    &

$\rm 24\,\mu m $    &
$\rm 70\,\mu m $    & 
$\rm 160\,\mu m $    &

$\rm 1.4\,GHz $    &

redshift \\

\hline
2MASS\_02 & $3.91\pm0.01$ & $4.10\pm0.01$ & $4.29\pm0.01$ & $4.40\pm0.01$ & $4.52\pm0.01$& $<4.57$& $<4.92$ & 5.6 & 0.635
\\

2MASS\_03 & $3.93\pm0.01$ & $4.04\pm0.01$ & $4.19\pm0.01$ & $4.28\pm0.01$ & $4.35\pm0.02$& $5.71\pm0.03$& $5.32\pm0.05$ & 3.6 & 0.433
\\

2MASS\_04 & $3.60\pm0.01$ & $3.74\pm0.01$ & $3.85\pm0.01$ & $4.01\pm0.01$ & $4.26\pm0.02$& $5.32\pm0.03$& $5.32\pm0.05$ & $-$ & 0.422
\\

2MASS\_05 & $3.27\pm0.01$ & $3.45\pm0.01$ & $3.74\pm0.01$ & $3.99\pm0.01$ & $4.36\pm0.01$& $5.28\pm0.03$& $5.12\pm0.05$ & $-$ & $3.10^{ph}$ 
\\

2MASS\_06 & $3.96\pm0.01$ & $4.09\pm0.01$ & $4.25\pm0.01$ & $4.36\pm0.01$ & $4.78\pm0.01$& $5.41\pm0.03$& $5.09\pm0.05$ & 8.9 & 0.264
\\

2MASS\_07 & $3.40\pm0.01$ & $3.62\pm0.01$ & $3.83\pm0.01$ & $4.03\pm0.01$ & $4.29\pm0.02$& $4.77\pm0.03$& $<4.51$ & $-$ & $2.73^{ph}$ 
\\

2MASS\_08 & $3.08\pm0.01$ & $3.26\pm0.01$ & $3.44\pm0.02$ & $3.74\pm0.01$ & $4.39\pm0.01$& $5.26\pm0.03$& $<4.85$ & $-$ & 2.160
\\

2MASS\_10 & $3.80\pm0.01$ & $3.98\pm0.01$ & $4.14\pm0.01$ & $4.28\pm0.01$ & $4.54\pm0.01$& $4.96\pm0.03$& $<4.95$ & 63.2 & 0.804
\\

2MASS\_11 & $3.53\pm0.01$ & $3.67\pm0.01$ & $3.91\pm0.01$ & $4.07\pm0.01$ & $4.18\pm0.02$& $<4.57$& $4.51\pm0.05$ & 919.7 & 2.210
\\

2MASS\_12 & $3.89\pm0.01$ & $4.02\pm0.01$ & $4.17\pm0.01$ & $4.29\pm0.01$ & $4.78\pm0.01$& $5.82\pm0.03$& $5.44\pm0.05$ & 0.8 & 0.291
\\
\hline
\end{tabular}\label{tab_obs2} 
\end{table*}

\section{Observations}\label{sec_observations}

In  addition to  the  available optical  (SDSS;  $ugriz$) and  near-IR
(2MASS;  $JHK$)   broad  band  data,  which  are   presented  in  Table
\ref{tab_obs1},  we have also  carried out  mid- and  far-IR photometry
with Spitzer as well as  optical and near-IR spectroscopy to determine
redshifts. These observations are described below.

\subsection{Mid- and far-Infrared photometry}

The  Spitzer  mid-  and  far-infrared  photometric  observations  were
performed during Cycle-2 as part  of the General Observer program with
identification  number 30607  (PI Georgakakis).   The mid-  and far-IR
flux  densities   of  the  target  sources  are   presented  in  Table
\ref{tab_obs2}.

The $\rm 3.6-8.0\mu  m$ data were taken with  the IRAC (Infrared Array
Camera)  between  July  9  and  December 28  2006.   Each  observation
consists of 5 separate integrations of 2\,s each organised in a random
Gaussian  dither pattern.   The median  separation  between successive
offsets is 0.9\,arcmin.  The observations were run through the S14.4.0
version of the Spitzer Science  Center pipeline described in IRAC data
handbook.  The basic calibration data (BCD) products are processed for
dark  current, bias offset,  linearisation, flat-fielding,  cosmic ray
detection and the flux calibration.   The individual frames of a given
source  were then  registered  and co-added  into  the final  mosaiced
image.  The uncertainty in the calibration factor applied to the final
mosaics is about  10 per cent.  Following the  IRAC data handbook flux
densities  are estimated by  using apertures  of 6\,arcsec  radius and
then applying aperture correction  factors to account for the extended
Spitzer point spread function (PSF) in different wavebands.  These are
estimated 1.06 for  IRAC 3.6, $\rm 4.5\,\mu m$ bands  and 1.10 for the
5.8  and $8.0\,\mu  m$ bands.  Colour corrections  are expected  to be
small as the mid-IR spectra of the sources in the sample approximate a 
power-law with  $\nu f\nu \rm  \approx constant$. We  therefore choose
not to apply colour corrections.

The 24, 70,  and $\rm 160 \mu m$ data were  taken with MIPS (Multiband
Imaging Photometer for Spitzer) between  July 14 2006 and September 15
2007. The observations used  the MIPS small-field photometry mode with
an exposure time of 3\,s for  individual frames. The 24 and $\rm 70\mu
m$  data  were obtained  in  one  cycle, while  the  $\rm  160 \mu  m$
observations  were  carried out  in  3  cycles.   The total  on-source
exposure times were  48, 38 and 25\,s  at 24, 70, and $\rm  160 \mu m$
respectively.  The MIPS images were created from raw data frames using
the MIPS  Data Analysis Tools  \citep[MIPS DAT;][]{Gordon2005} version
3.10 along with additional processing steps.  The processing steps for
the 70  and $\rm 160 \mu  m$ data are  similar, but the steps  for the
24\,$\rm \mu m$ data are  significantly different from these other two
bands.   The  24\,$\rm  \mu  m$  data processing  is  described  first
followed  by  descriptions  of  the  70  and  $\rm  160  \mu  m$  data
processing.

The individual 24\,$\mu$m frames  were first processed through a droop
correction  (removing   an  excess  signal  in  each   pixel  that  is
proportional to the signal in the entire array) and were corrected for
non-linearity  in the ramps.   The dark  current was  then subtracted.
Next, scan-mirror-position  dependent flats were applied  to the data,
latent images were removed, and scan-mirror-position independent flats
were  applied to  the data.   Following this,  the zodiacal  light was
measured by  fitting planes to  the background regions in  each frame,
and these planes  were then subtracted from the  data.  Next, a robust
statistical analysis  was applied  to cospatial pixels  from different
frames in  which statistical outliers (e.g. pixels  affected by cosmic
rays) were masked out.  After this, final mosaics were made with pixel
sizes of $1.5^{\prime\prime}$, residual backgrounds in the images were
subtracted, and the data were calibrated into astronomical units.  The
calibration  factor  for  the  24\,$\rm   \mu  m$  data  is  given  by
\citet{Engelbracht2007}   as   $(4.54\pm0.18)   \times  10^{-2}$   MJy
sr$^{-1}$  [MIPS  instrumental  unit]$^{-1}$.  A  10.5\,arcsec  radius
aperture  is used  for  the photometry.   The  correction factor  that
accounts for PSF  losses is estimated to be  1.22.  The mid-IR spectra
of the sample sources is close  to a power-law.  For this SED type the
colour correction at the 24\,$\rm \mu m$ band is small ($<$3 per cent)
and we choose not to apply it.

In the 70 and  $\rm 160 \mu m$ data processing, the  first step was to
fit ramps  to the reads to  derive slopes, during  which readout jumps
and cosmic ray  hits were also removed and  an electronic nonlinearity
correction  was applied.   Next, the  stim flash  frames taken  by the
instrument were  used as  responsivity corrections.  The  dark current
was subtracted from the  data, an illumination correction was applied,
and  short term variations  in the  the signal  (often referred  to as
drift) were removed.  Next,  a robust statistical analysis was applied
to  cospatial  pixels  from  different  frames  in  which  statistical
outliers (e.g. pixels affected by  cosmic rays) were masked out.  Once
this  was  done,  final  mosaics  were made  using  square  pixels  of
4.5\,arcsec  for  the  $\rm  70  \mu  m$ data  and  9\,arcsec  for  the
160\,$\mu$m  data.    The  residual  backgrounds   were  measured  and
subtracted  from the  images. Finally,  flux calibration  factors were
applied to the data.  The $\rm  70 \mu m$ calibration factors given by
\citet{Gordon2007} are  $702 \pm 35$ MJy  sr$^{-1}$ [MIPS instrumental
unit]$^{-1}$, and the  $\rm 160 \mu m$ calibration  factor is given by
\citet{Stansberry2007} as $41.7\pm5$  MJy sr$^{-1}$ [MIPS instrumental
unit]$^{-1}$. For the  photometry we use apertures with  radii 4.5 and
9\,arcsec at  70 and $\rm 160  \mu m$ respectively. In  this case, the
correction  factors that  account for  PSF  losses are  1.30 and  1.87
respectively. Colour corrections  for the MIPS 70 and  $\rm 160 \mu m$
bands are typically smaller than the calibration uncertainties and are
therefore not  applied to  the data. In  the case of  non-detection we
assign an upper limit to the flux density which corresponds to 5 times
the standard deviation of the background.

\subsection{Spectroscopy}

Optical spectroscopy  for the  target sources was  carried out  at the
Kitt  Peak National  Observatory  (KPNO) 4-m  telescope  and the  4.2m
William Herschel Telescope (WHT).

The  KPNO  observations  used  the  Ritchey-Chretien  spectrograph  in
single-slit mode with the BL\,181  grism blazed at 7500$\rm \AA$. This
setup provides  a resolution of  about 7\AA\, in the  wavelength range
$\rm  5500-9500\AA$. An  exposure time  of 20min  was  adopted.  These
observations were carried out in March 10 2006.

The   WHT  spectroscopy   used  the   ISIS   (Intermediate  dispersion
Spectrograph  and  Imaging System)  double  armed spectrograph  during
service time  in June 17 and  July 8 2007. The  observations were made
using the 5300\AA\,  dichroic and the R300B and  R158R gratings in the
blue  and red  arms  of the  spectrograph  respectively. The  spectral
resolution was  about 7\AA\, for the  red arm and 4\AA\,  for the blue
arm spectrum.  The total  on-source integration time was 20\,min split
into two 10\,min exposures.

The data were reduced following standard methods as implemented in the
{\sc noao} package of {\sc  iraf}.  After subtraction of the bias, the
data  were  flat-fielded using  observations  of internal  calibration
lamps. Cosmic ray  events were removed using the  Laplacian Cosmic Ray
Identification  package \citep[{\sc  lacosmic};][]{vanDokkum2001}. The
one dimensional spectra were extracted and wavelength calibrated using
observations of  CuNe and  CuAr arc lamps.   For the  flux calibration
observations of the spectrophotometric standard stars BD 282411 and BD
332642 were used.  Because of the apparent faintness of our targets at
wavelengths shorter  than 5000\AA\, the WHT/ISIS blue  arm spectra did
not show any signal and are therefore not presented in this paper.

Near-infrared spectroscopy  of SDSS/2MASS  EROs was obtained  in queue
mode on the UK Infrared  Telescope (UKIRT).  The observations used the
UKIRT Imaging SpecTrometer (UIST) in  long slit mode with a mixture of
the IJ  and HK grisms, the  exact selection depending  on the expected
redshift of the object  based on photometric redshift estimation.  The
observations were carried out in  clear conditions on 18, 22, 23 June,
2007.  Integration  times were $\sim$ 1.5  hours for the  HK grism and
$\sim$ 45  minutes for the IJ  grism. The data were  reduced using the
automated  ORACDR  system which  performed  flat fielding,  wavelength
calibration based  on arc spectra,  subtracted the chopped  images and
coadded separate sub-exposures. Flux calibration was carried out using
the  BS5019 and  BS8 spectrophotometric  standards. No  correction for
slit  losses were  applied so  the flux  calibration is  correct  in a
relative sense. The spectra were  then analyzed using a combination of
IRAF and IDL  routines to determine redshifts and  to extract relevant
emission line parameters.

\begin{figure*}
\begin{center}
\rotatebox{0}{\includegraphics[height=0.9\columnwidth, angle=270]{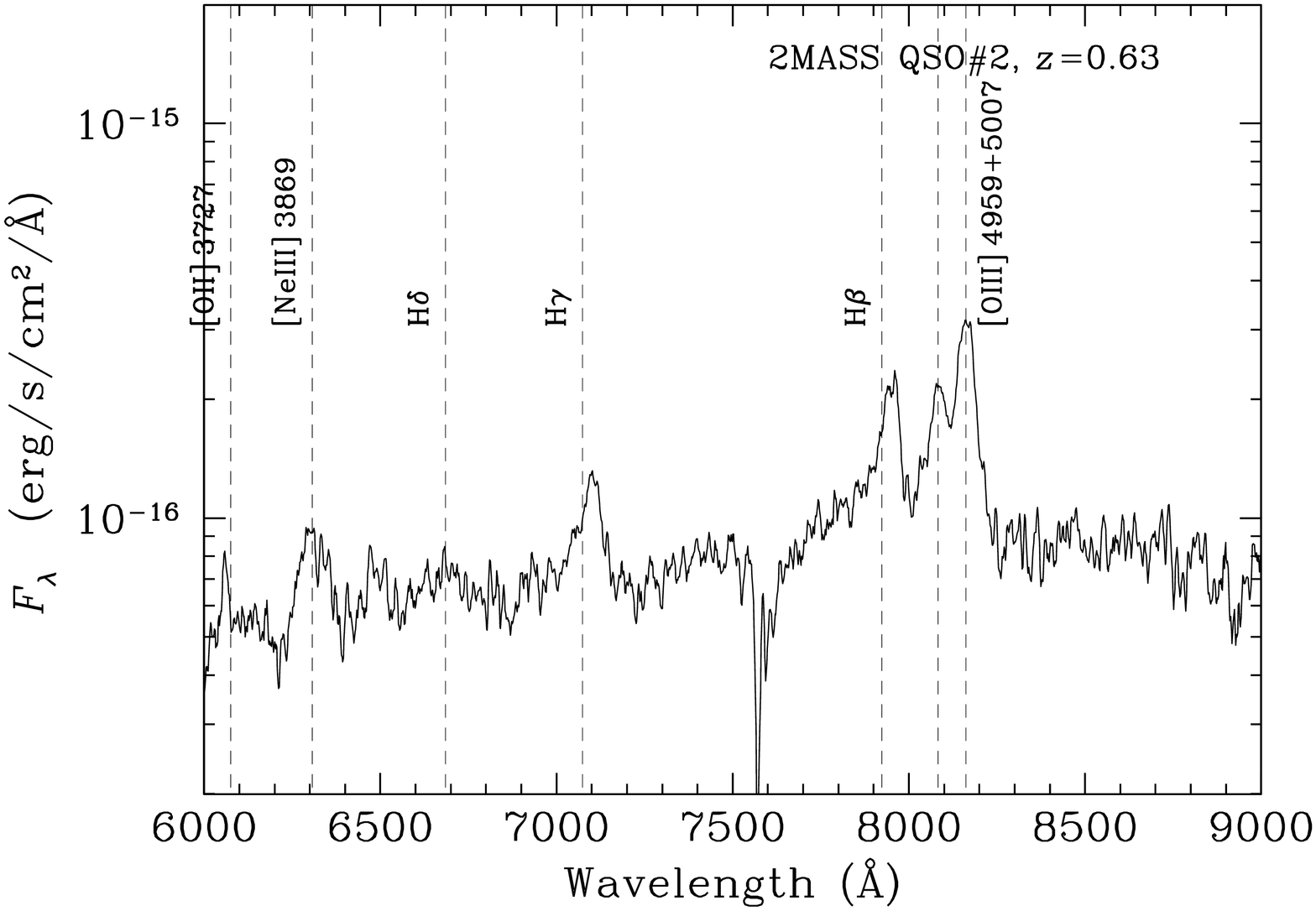}}
\rotatebox{0}{\includegraphics[height=0.9\columnwidth, angle=270]{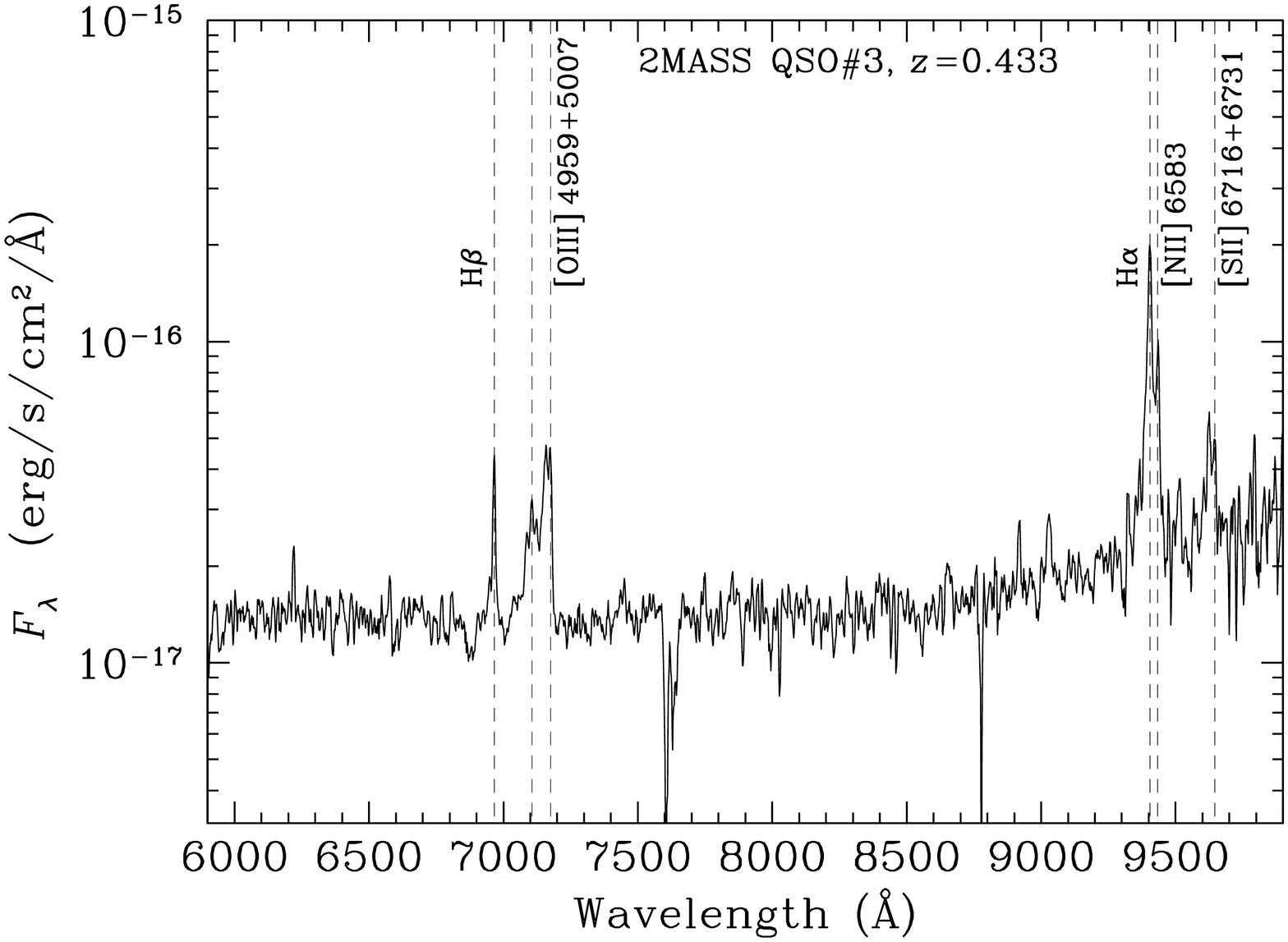}}
\rotatebox{0}{\includegraphics[height=0.9\columnwidth, angle=270]{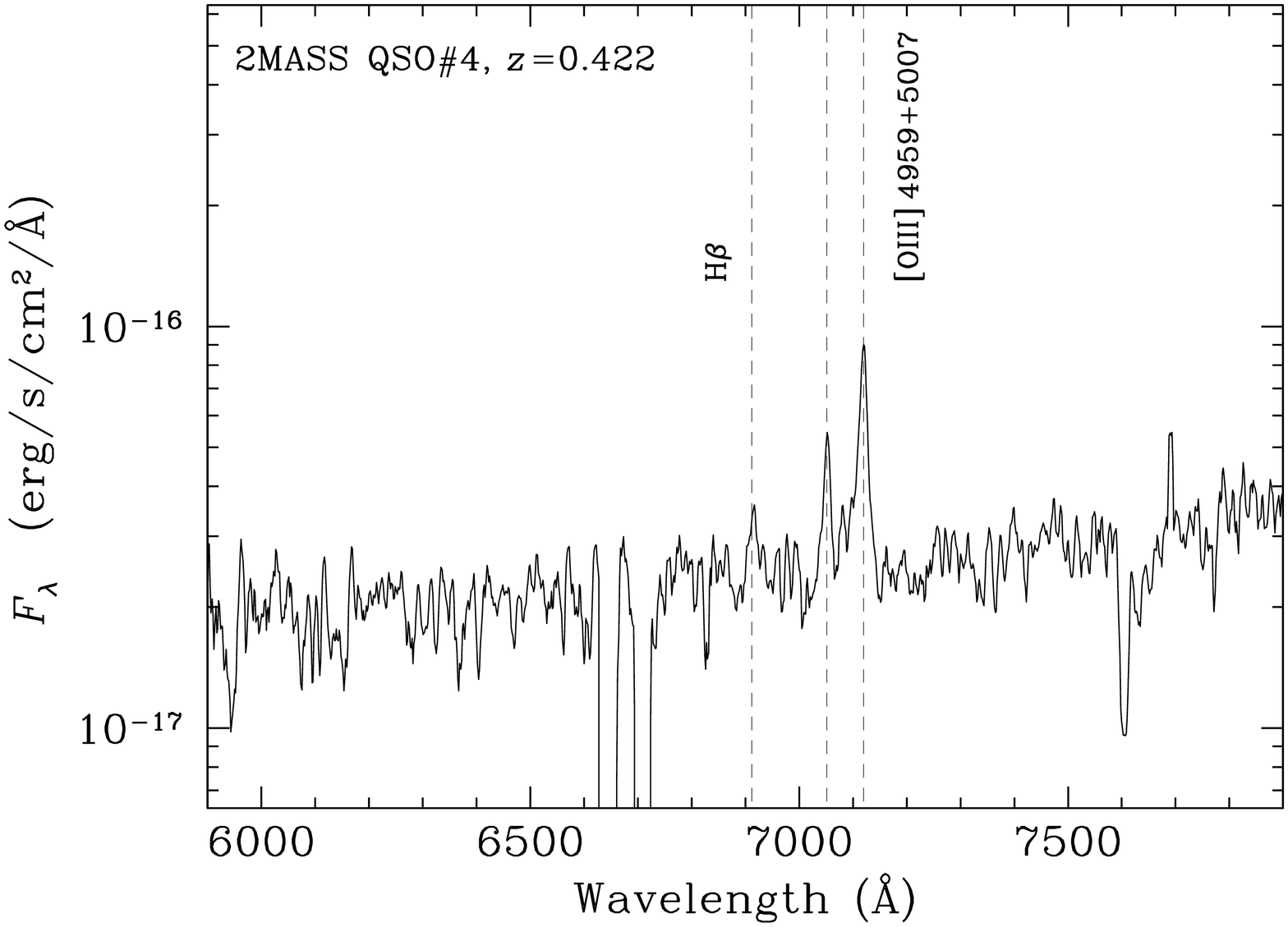}}
\rotatebox{0}{\includegraphics[height=0.9\columnwidth, angle=270]{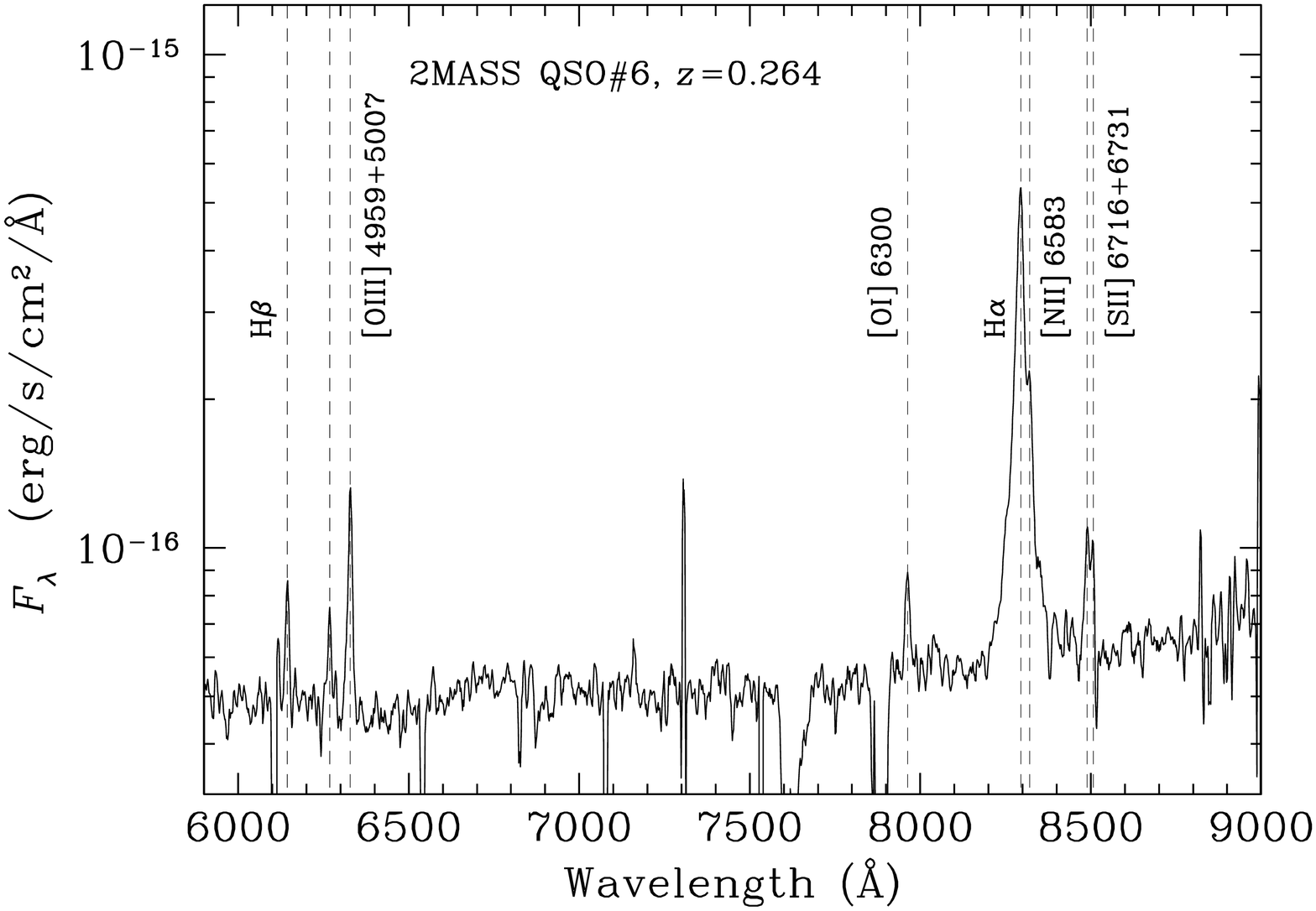}}
\rotatebox{0}{\includegraphics[height=0.9\columnwidth, angle=270]{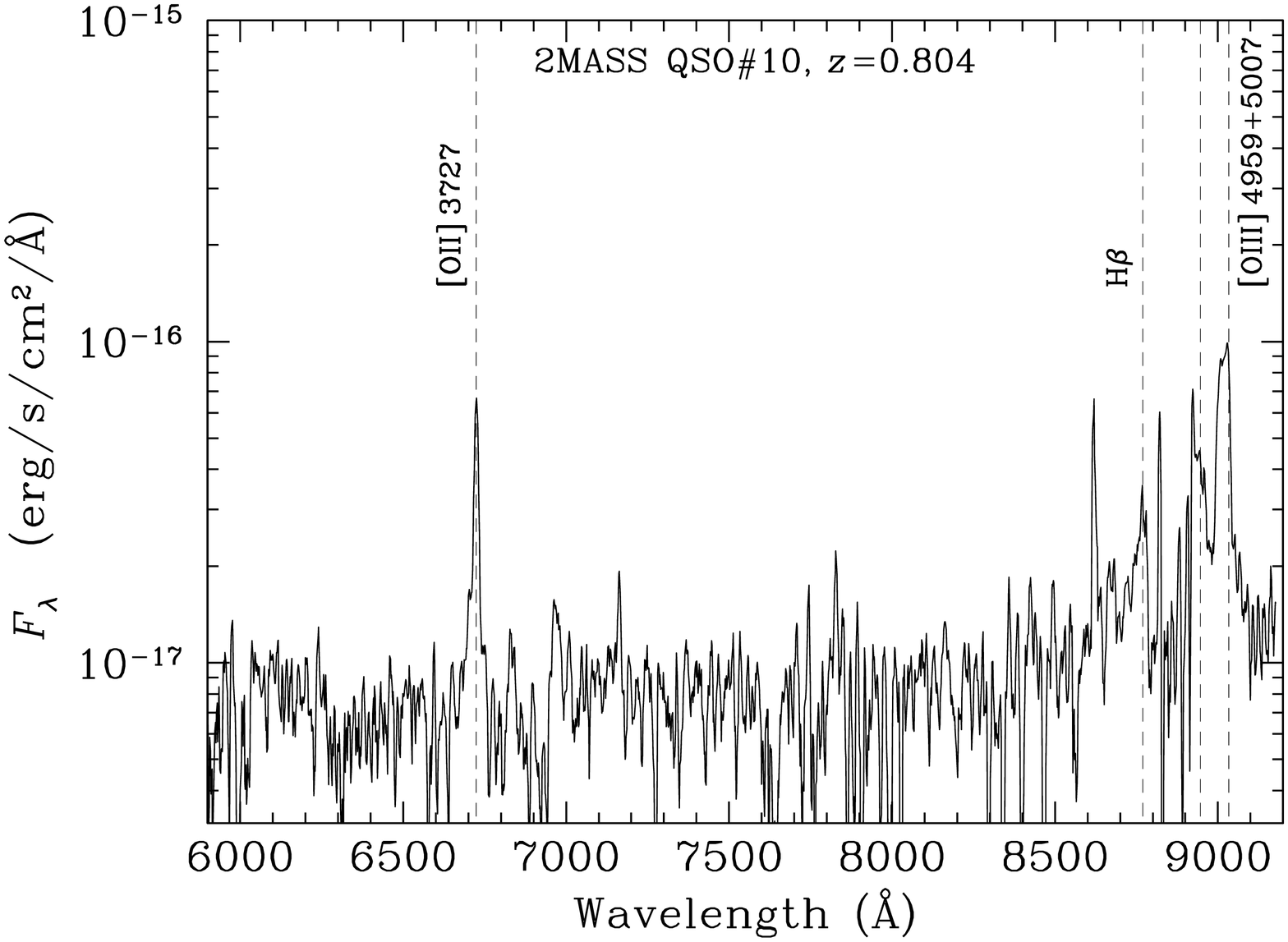}}
\rotatebox{0}{\includegraphics[height=0.9\columnwidth, angle=270]{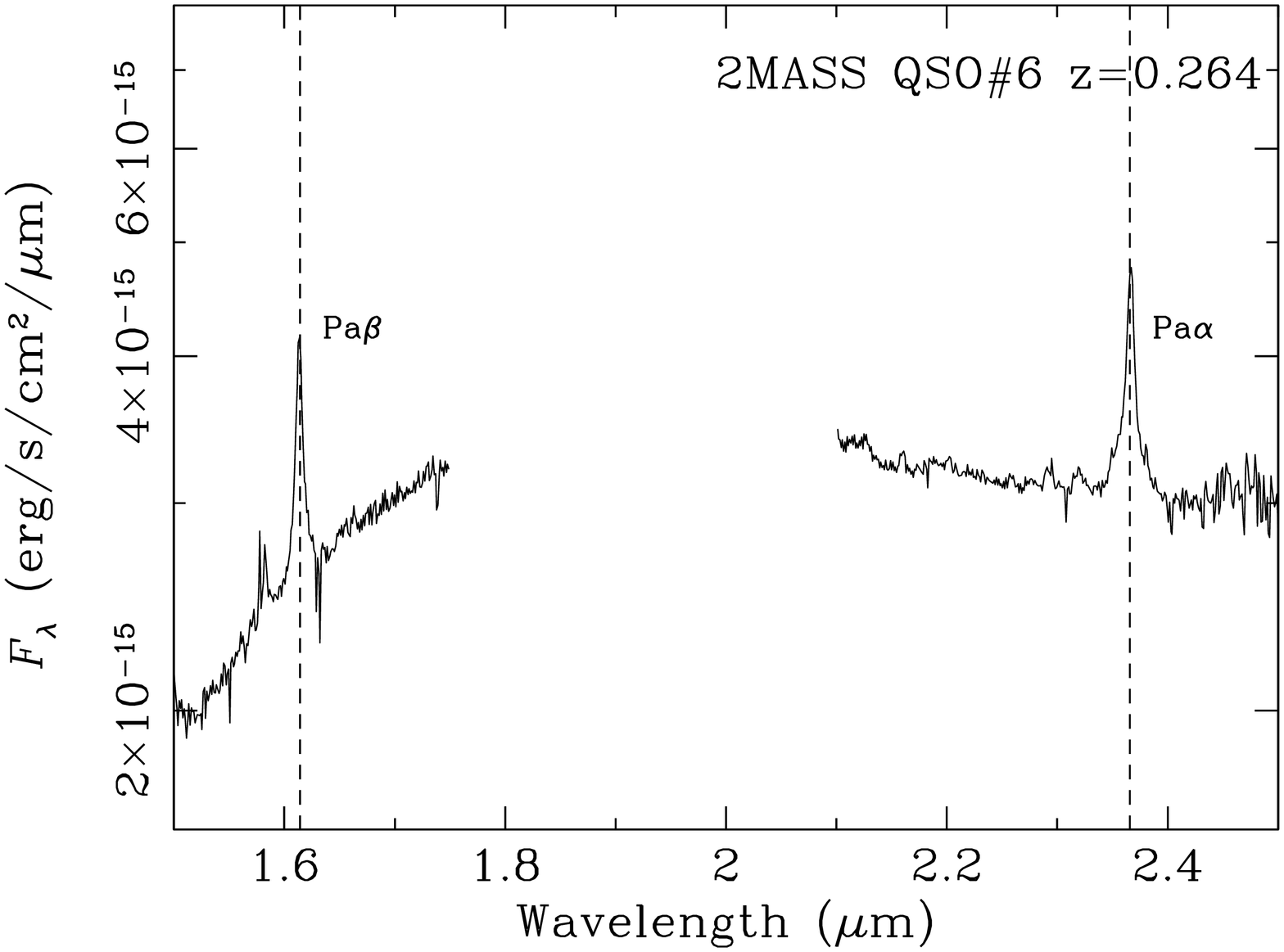}}
\rotatebox{0}{\includegraphics[height=0.9\columnwidth, angle=270]{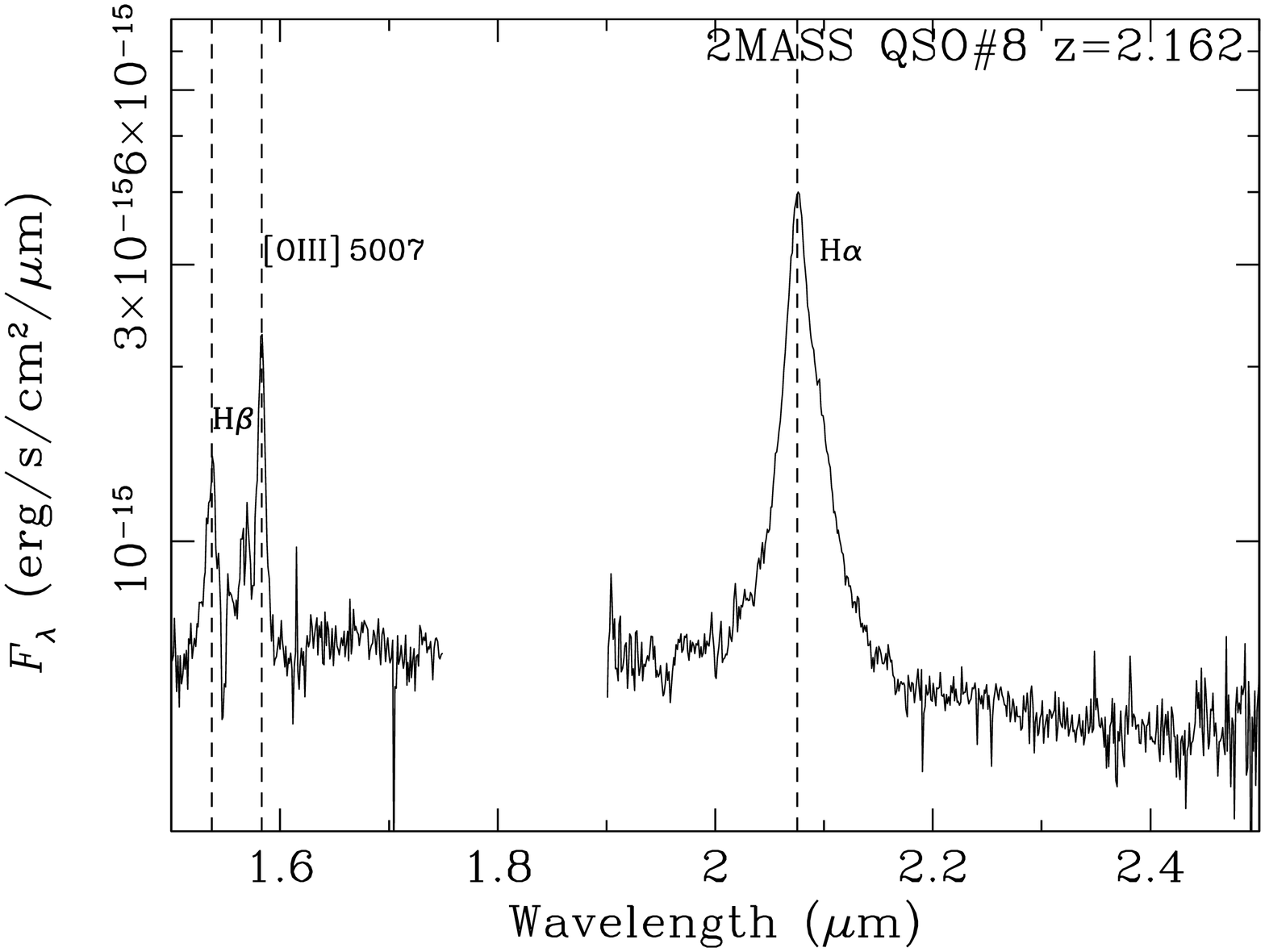}}
\rotatebox{0}{\includegraphics[height=0.9\columnwidth, angle=270]{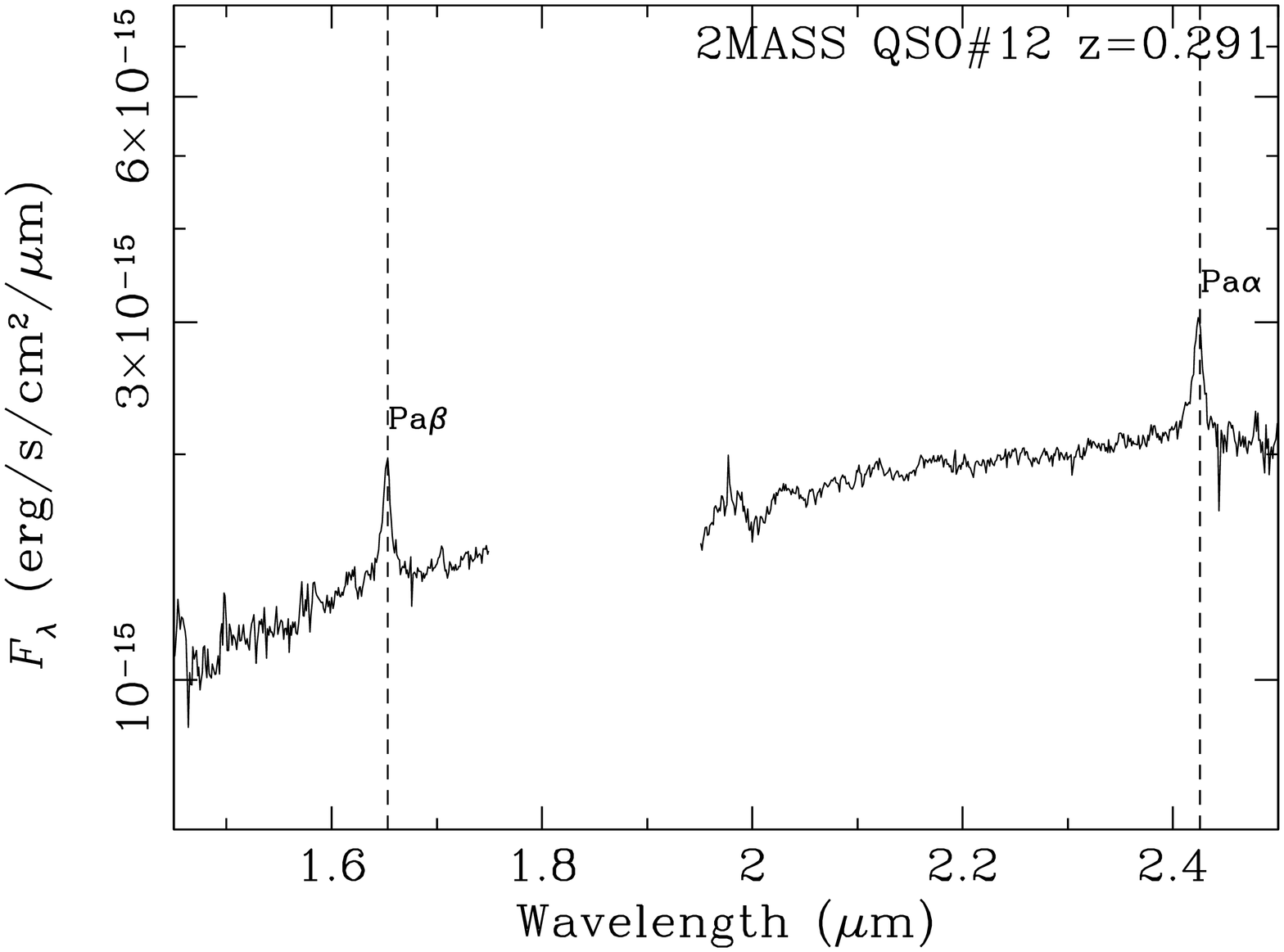}}
\end{center}
\caption{Optical and  near-IR spectra  of the sources  with successful
redshift  determination from our  own spectroscopic  observations. The
gap between  near-IR spectra corresponds the  gap between the  H and K
windows of the HK grism. }\label{fig_spec}
\end{figure*}

Spectroscopic redshifts  are available for 8/10 2MASS  sources, 7 from
our  own  observations  and  1   from  the  literature.  The  lack  of
spectroscopic redshift  determinations for 2  sources in the  sample is
because of low  S/N optical spectra.  The optical  and near-IR spectra
of the  sources with successful  redshift determinations are  shown in
Figure   \ref{fig_spec}.   The  redshifts   are  presented   in  Table
\ref{tab_obs2} and estimates of the  FWHM (Full Width Half Maximum) of
emission lines are  discussed in the Appendix and  are listed in Table
\ref{tab_restframe} .

\section{Spectral Energy Distributions}\label{sec_sed}

The observed  optical to mid-IR Spectral Energy  Distribution (SED) of
the sample  sources are modeled following the  methods fully described
in \cite{rowan2005,rowan2008}. In brief  the $U$-band to $\rm 4.5\,\mu
m$ photometric data  are fit using a library  of 8 templates described
by \cite{Babbedge2004}, 6 galaxies (E,  Sab, Sbc, Scd, Sdm and sb) and
2 AGN.  At longer  wavelengths ($\rm 5.8 - 24 \, \mu  m$) any dust may
significantly  contribute  or  even  dominate the  observed  emission.
Before fitting models to these wavelengths the stellar contribution is
subtracted  from the  photometric data  by extrapolating  the best-fit
galaxy template  from the previous  step.  The residuals are  then fit
with a  mixture of four templates:  cirrus \citep{Efstathiou2003}, AGN
dust  tori  \citep{rowan1995,   Efstathiou1995},  M\,82  and  Arp\,220
starbursts  \citep{Efstathiou2000}. The  modeling above  provides both
information on the dominant emission  mechanism in the optical and the
infrared (AGN vs star-formation) and an estimate of the total infrared
luminosity, $L_{TOT}$,  of our sample sources in  the wavelength range
$\rm  3-1000\mu m$.   As  discussed by  \cite{rowan2005} $L_{TOT}$  is
expected to be accurate within a  factor of two.  In this exercise the
redshift  is  fixed  to  the spectroscopically  determined  value,  if
available (8/10 sources).   For the two sources in  the sample without
spectroscopic  redshifts  we   also  estimate  photometric  redshifts,
although we caution  that these are likely to  be uncertain because of
their extreme optical/near-IR  colours.  From the spectroscopic sample
we estimate  that the fraction  of catastrophic redshifts,  defined as
those  with $(z_{spec}-z_{phot})/(1+z_{spec})>0.15$,  is  38 per  cent
(3/8).        The      rms       value      of       the      quantity
$(z_{spec}-z_{phot})/(1+z_{spec})$,   after   excluding  catastrophic
failures,  is 0.07 and  provides an  estimate of  the accuracy  of the
photometric redshifts.

\begin{figure*}
\begin{center}
 \rotatebox{0}{\includegraphics[height=1.9\columnwidth]{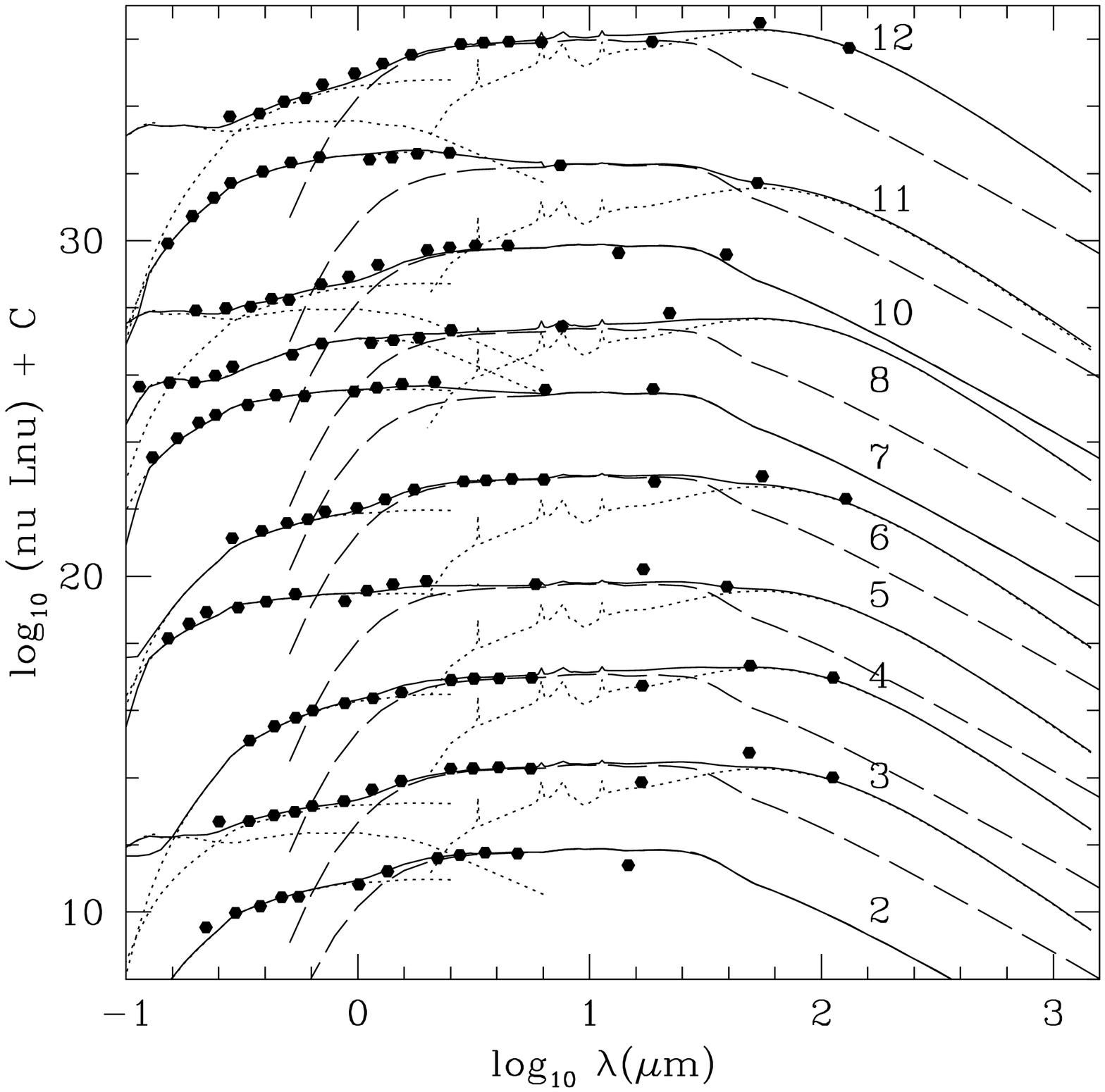}} 
\end{center}
\caption{Observed  Spectral  Energy  Distributions  of the  red  2MASS
sources. The  dots are the observed UV--to--far-IR  photometry and the
continuous  lines  are the  best-fit  models.   Different sources  are
labeled  and are  offset by  an  arbitrary constant  for clarity.  The
mid-IR  wavelengths  of  all sources  are  fit  by  a hot  dust  torus
component  (dashed line)  which dominates  at these  part of  the SED.
Some sources  also require  an additional starburst  component (dotted
line at  infrared wavelengths)  to fit the  far-IR data.   The optical
part of the  SED is fit with an reddened  QSO template.  The reddening
is  listed  in  Table  \ref{tab_restframe}. Some  sources  require  an
additional component  to fit the  $u$ and/or $g$-band  SDSS photometry
(2MASS\_3, 2MASS\_10, 2MASS\_12).   A star-formation component is used
here,  although we  cannot exclude  the possibility  of  scattered AGN
light. }\label{fig_sed}
\end{figure*}

Figure  \ref{fig_sed} overplots  the best-fit  models to  the observed
SEDs of the  red 2MASS sources.  The derived  parameters are presented
in  Table \ref{tab_restframe}.   In  summary, the  mid-IR  SED of  all
sources  is dominated  by  hot dust,  which  is fit  by  an AGN  torus
component.  An  additional starburst template  is required to  fit the
far-IR data of some  sources in Table \ref{tab_restframe}. The optical
part of the  SED is fit with a reddened QSO  template adopting the SMC
extinction   curve  of   \cite{Richards2003}.   The   derived  optical
extinctions are in the range $A_V=1.3-3.2$ ($E(B-V) \approx 0.4-1.1$).

\begin{table*} 

\caption{The rest-frame properties of the 2MASS QSOs. The columns are: 
1: name; 
2: total infrared luminosity in solar units ($L_\odot=\rm
3.83\times10^{33} \, erg \, s^{-1}$) and in the wavelength interval $\rm
3-1000\mu m$. This is estimated from the SED fits described in section
\ref{sec_sed}; 
3: Infrared templates used to fit the mid- and far-IR
observations. There are sources where both an AGN torus and a starburst
component are required.  
4: Ratio of the total infrared luminosity
in the dust torus ($3 - 1000 \mu m$) to the reddening corrected optical ($0.1 - 3\mu m$) QSO luminosity. This ratio has been interpreted as a measure of the dust
covering factor in AGN (Rowan-Robinson et al. 2008).
5: Optical $V$-band extinction in magnitudes estimated from fitting
the observed SED as described in section \ref{sec_sed};
6: $Ks$-band absolute magnitude. The $Ks$-band flux density is estimated
directly from the observations by simply convolving the rest-frame SED
of each source (i.e. corrected for redshift) with the $Ks$-band filter
function.  
7: Radio power at 1.4\,GHz in units of W/Hz. A radio spectrum of the
form $f_\nu \propto \nu^{-0.75}$ was adopted for the calculation of
the k-correction.  
8: FWHM in km/s of the broad component of the observed emission
lines. We do not list FWHM for sources with narrow emission lines only
(e.g. 2MASS\_04) or without optical spectroscopy from our own
follow-up program. The line used to measure the FWHM is also listed in
brackets.}
\centering
\scriptsize
\begin{tabular}{l cccc cc c}
\hline 
ID &

$\log L_{TOT}$    & 
IR template    &
$\log \frac{L_{torus}}{L_{opt}}$ & 
$A_V$    & 
$M_{K_s}$    &

$\log L_{1.4} $    &
FWHM   \\

 &

($L_{\odot}$) & 
    &
    & 
(mag) & 
(mag) &

(W/Hz)&
(km/s)   \\

\hline
2MASS\_02 & 13.11 & torus   & 0.15    & 2.3  & -29.49 & 24.94 &  2000 ($\rm H\beta$)\\ 
\\

\multirow{2}{*}{2MASS\_03} & 12.57 & torus  & \multirow{2}{*}{0.40}  & \multirow{2}{*}{1.9}  & \multirow{2}{*}{-28.48} & \multirow{2}{*}{24.36} &  \multirow{2}{*}{1500 ($\rm H\beta$)}\\
                           & 12.37 & M\,82  &                       &                       &                         &                        &  \\
\\

\multirow{2}{*}{2MASS\_04} & 12.27 & torus  & \multirow{2}{*}{-0.20} & \multirow{2}{*}{3.2}  & \multirow{2}{*}{-27.56}  & \multirow{2}{*}{--}    & \multirow{2}{*}{--} \\
                           & 12.37 & M\,82  &                       &                       &                          &                        &  \\
\\

\multirow{2}{*}{2MASS\_05} & 14.17 & torus & \multirow{2}{*}{-0.20} & \multirow{2}{*}{1.3}  & \multirow{2}{*}{-32.15}  & \multirow{2}{*}{--}    &  \multirow{2}{*}{--}\\
                           & 13.87 & M\,82 &                        &                       &                          &                        &  \\
\\

\multirow{2}{*}{2MASS\_06} & 12.12 & torus  & \multirow{2}{*}{0.25} & \multirow{2}{*}{2.3}  & \multirow{2}{*}{-27.34}  & \multirow{2}{*}{24.25} &  \multirow{2}{*}{1500 ($\rm Pa\,\alpha$)}\\
                           & 11.72 & M\,82  &                        &                       &                          &                        &  \\
\\

2MASS\_07& 13.86 & torus     & -0.80  & 1.5  & -31.98 & -- &  -- \\ 
\\

\multirow{2}{*}{2MASS\_08} & 13.75 & torus  & \multirow{2}{*}{0.00}   & \multirow{2}{*}{1.2}  & \multirow{2}{*}{-30.64}  & \multirow{2}{*}{--} & \multirow{2}{*}{4700 ($\rm H\alpha$)} \\
                           & 13.95 & M\,82  &                        &                       &                          &                     &  \\
\\

2MASS\_10 & 13.14 & torus      & 0.40  & 2.3  & -29.83 & 26.24 &  -- \\ 
\\

\multirow{2}{*}{2MASS\_11} & 13.65 & torus  & \multirow{2}{*}{-1.10}  & \multirow{2}{*}{2.3}  & \multirow{2}{*}{-31.69}  & \multirow{2}{*}{28.43} &  -- \\
                           & 12.85 & M\,82  &                       &                       &                          &                        &  \\
\\

\multirow{2}{*}{2MASS\_12} & 12.13 & torus   & \multirow{2}{*}{0.40} & \multirow{2}{*}{3.0}  &  \multirow{2}{*}{-27.36} & \multirow{2}{*}{23.32} &   \multirow{2}{*}{1600 ($\rm Pa\,\alpha$)}\\
                           & 12.33 & M\,82  &                         &                       &                           &                        &  \\

\hline
\end{tabular}\label{tab_restframe}
\end{table*}

\begin{figure*}
\begin{center}
\rotatebox{0}{\includegraphics[height=0.9\columnwidth, angle=270]{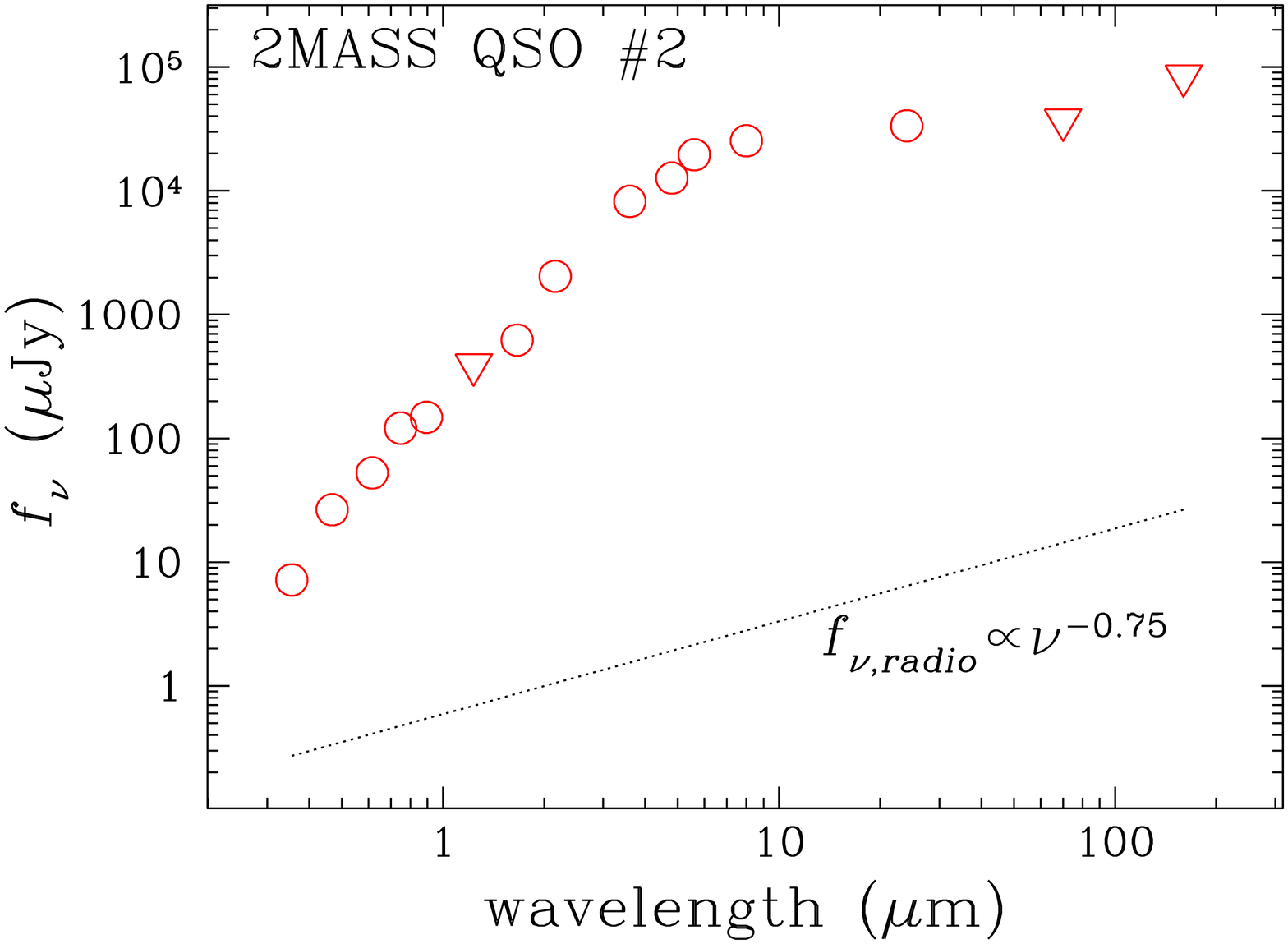}}
\rotatebox{0}{\includegraphics[height=0.9\columnwidth, angle=270]{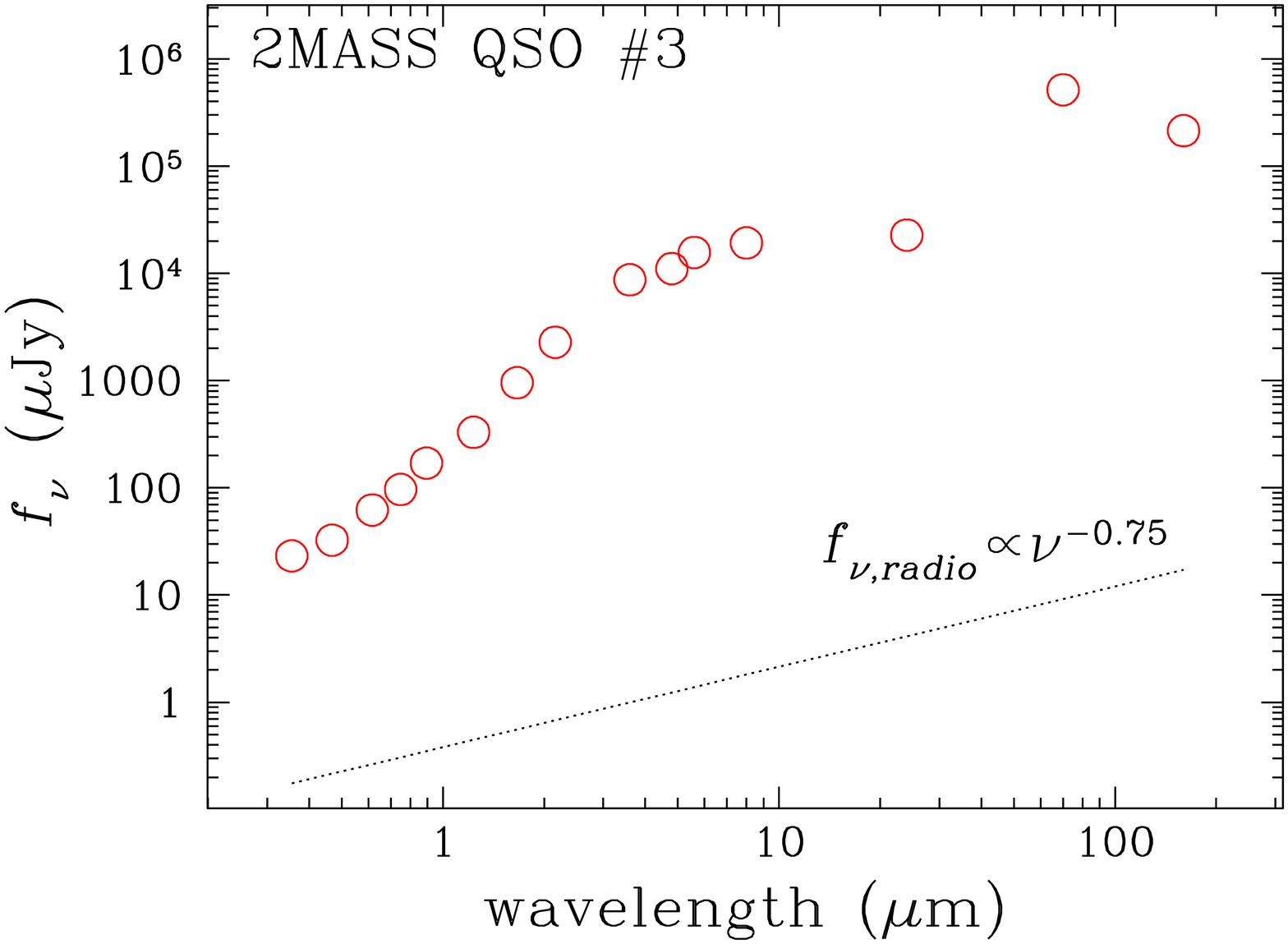}}
\rotatebox{0}{\includegraphics[height=0.9\columnwidth, angle=270]{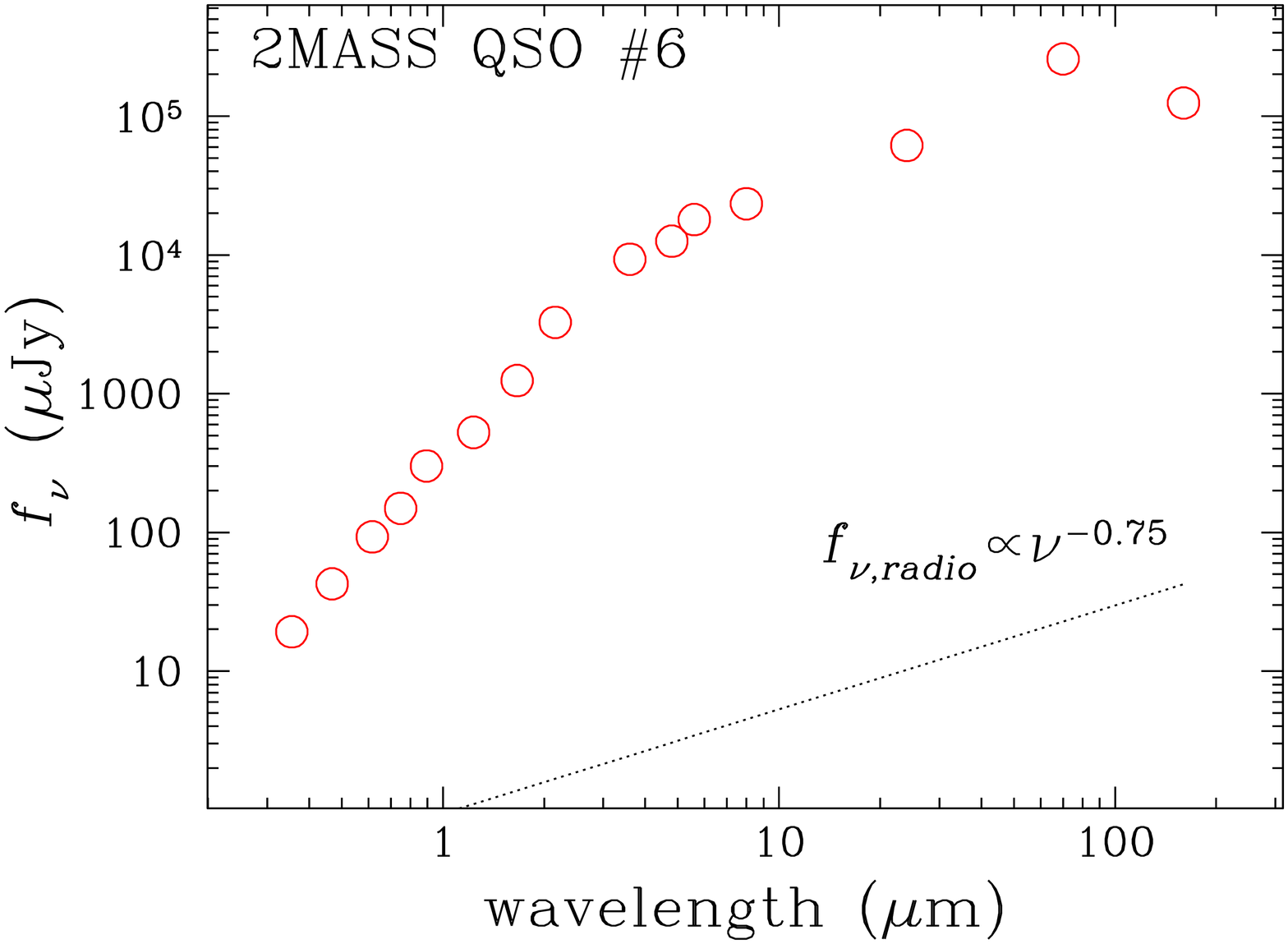}}
\rotatebox{0}{\includegraphics[height=0.9\columnwidth, angle=270]{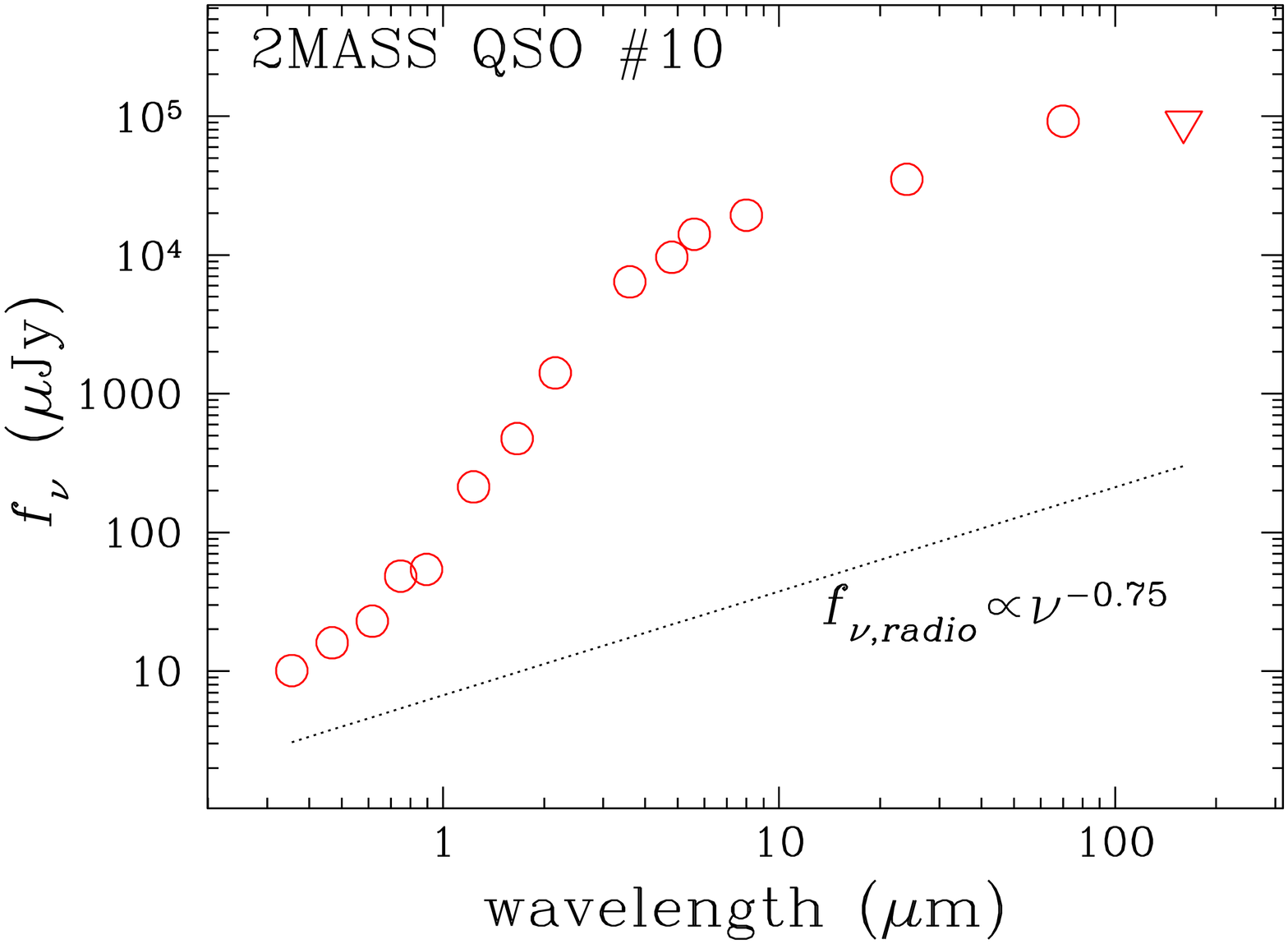}}
\rotatebox{0}{\includegraphics[height=0.9\columnwidth, angle=270]{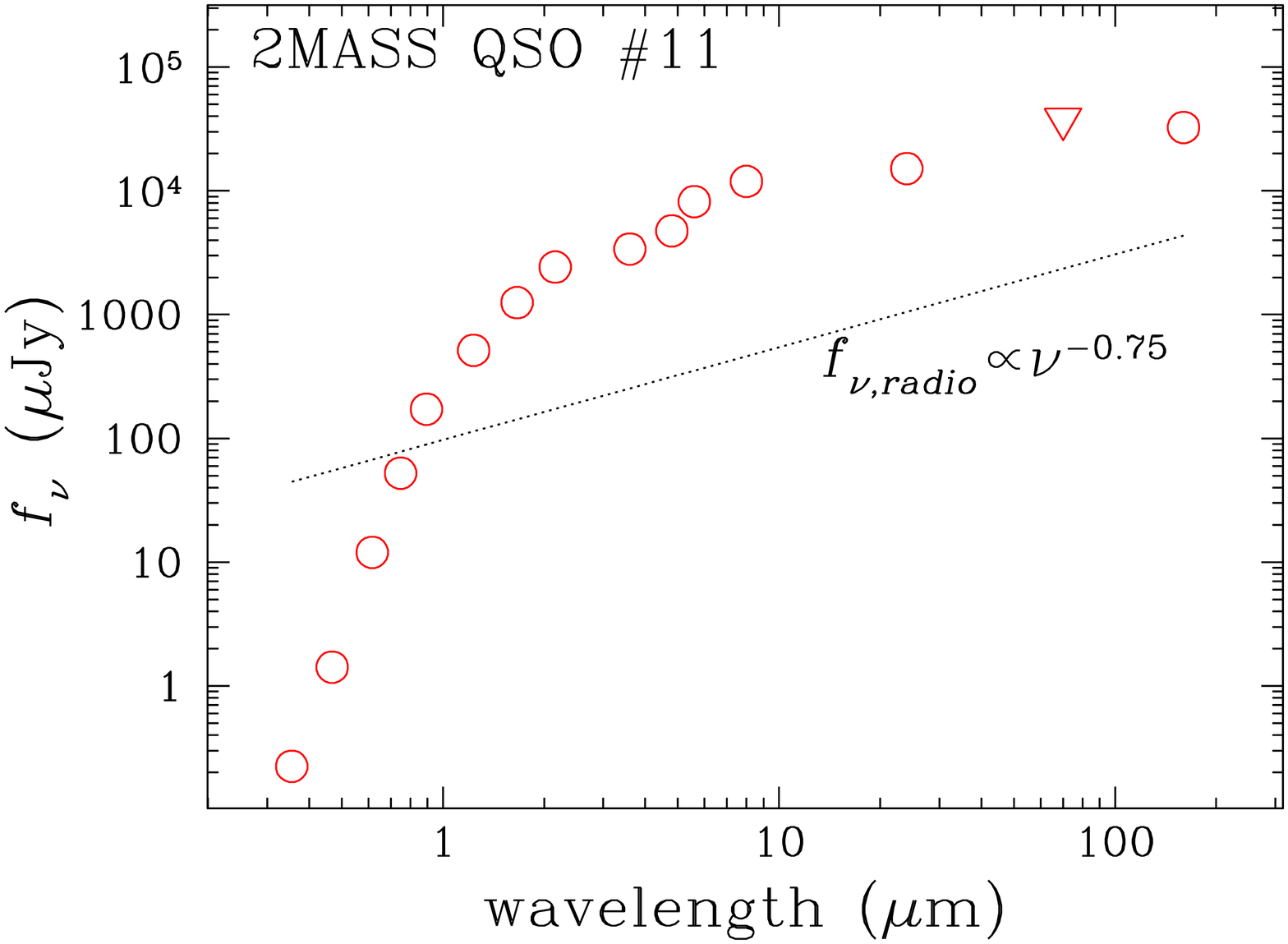}}
\rotatebox{0}{\includegraphics[height=0.9\columnwidth, angle=270]{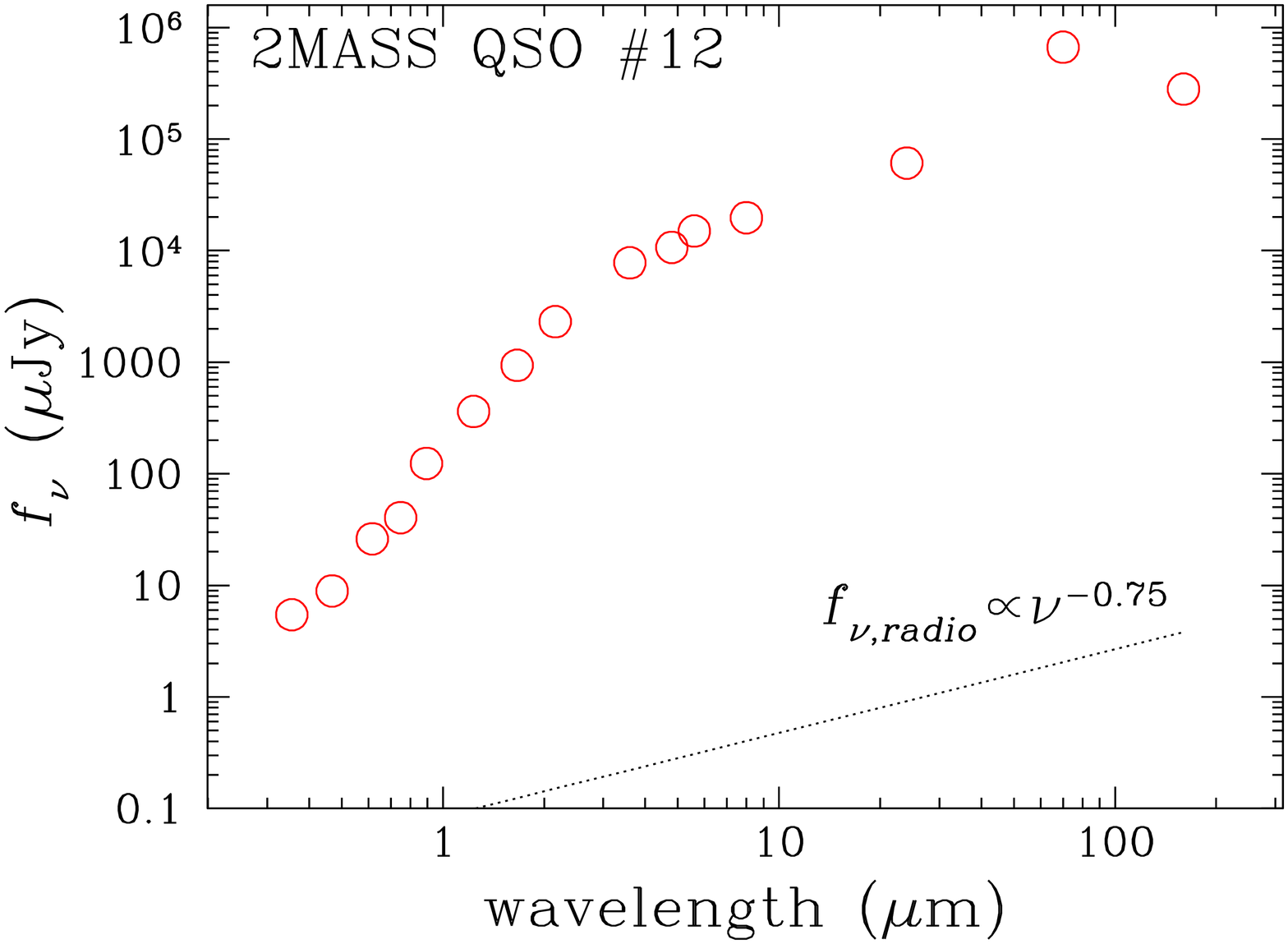}}
\end{center}
\caption{Spectral Energy  distributions of  the 2MASS QSOs  with radio
  counterparts in the FIRST or  the NVSS.  The dotted lines correspond
  to  the  radio  flux  density extrapolated  to  optical  wavelengths
  assuming a  radio spectrum of the form  $f_\nu \propto \nu^{-0.75}$,
  i.e. typical of  synchrotron emission \citep{Condon1992}.  The radio
  flux densities  are too faint to affect  the optical/near-IR colours
  of the  2MASS QSOs.   The only exception  is 2MASS\_11 with  a radio
  flux density  of about 1\,Jy.  Synchrotron emission  with a turnover
  frequency    in    the    UV/optical    part   of    the    spectrum
  \citep[e.g.][]{Whiting2001} could be responsible for the red colours
  of  this source.  We  note  however, that  the  possibility of  dust
  reddening cannot be ruled out.  For example, previous studies on the
  nature of  this source have  argued for dust associated  with either
  the   source  or   the  lens   \citep[e.g.][]{Hall2002,  Gregg2002}.
}\label{fig_radiosed}
\end{figure*}

\section{Nature of the 2MASS selected sample}\label{results}

The  nature  of  the  red  2MASS  sources  is  explored  by  combining
information  from the optical/near-IR  spectroscopy and  the broadband
UV--to--far-IR photometry.  There is  strong evidence that all sources
in the sample are powered by accretion on a central supermassive black
hole.  Firstly, for  seven out of the eight  sources for which optical
and/or near-IR spectra are available, either from our own observations
or from  the literature,  broad Balmer or  Paschen emission  lines are
found   with  widths   $>1000$\,km/s.    For  more   details  on   the
spectroscopic  properties  of  individual  objects  see  the  Appendix
section.   Additionally, the  red 2MASS  sources are  luminous  in the
mid-IR and their SEDs at these wavelengths are fit by the hot dust AGN
tori models of  \cite{rowan1995} and \cite{Efstathiou1995}.  The total
IR luminosity  ($\rm 3-1000\mu m$)  of this component is  estimated in
the range $\rm 10^{12}-10^{14}\, L_{\odot}$ ($\rm 4 \times 10^{45} - 4
\times 10^{47}\,  \rm erg \,  s^{-1}$), i.e.  exceeding the  limit for
either   ULIRGs   (Ultra-Luminous   Infrared  Galaxies)   or   HyLIRGs
(Hyper-Luminous  Infrared  Galaxies).    Adopting  an  AGN  bolometric
correction       factor       of       $L_{bol}/L_{IR}\approx       3$
\citep[e.g][]{Risaliti_Elvis2004},   the  luminosity   interval  above
exceeds the limit $L_{bol}\ga \rm 10^{46}\,erg/s$, which is often used
to differentiate  luminous QSOs from Seyfert  galaxies.  Therefore, in
the  following   we  refer   to  the  sample   of  sources   in  Table
\ref{tab_obs1} as 2MASS or red QSOs.  

The red  optical/near-IR colours of the  2MASS QSOs in Table  1 can be
attributed   to  either   an  intrinsically   red   optical  continuum
\citep[e.g.][]  {Richards2003}, synchrotron  emission with  a turnover
frequency    in    the     UV/optical    part    of    the    spectrum
\citep[e.g.][]{Whiting2001}   or  dust   reddening.   Each   of  these
possibilities is discussed below.

Radio selected  QSOs from the Parkes  half-Jansky flat-spectrum sample
\citep{Drinkwater1997}  have  a large  spread  in the  optical/near-IR
colours, with the  reddest objects having $B-K>7$ \citep{Francis2000}.
\cite{Whiting2001} proposed that  synchrotron emission with a turnover
frequency in the range $\rm 0.1-3\mu m$ can explain the red colours of
a sizeable fraction  of this population (about 40  per cent).  This is
relevant to  this study  as many  2MASS QSOs (6/10;  see Table  1) are
detected in the  NVSS \citep[][limit $\approx2.5$\,mJy]{Condon1998} or
the FIRST \citep[][limit $\approx1$\,mJy]{Becker1995} radio (1.4\,GHz)
surveys    with    luminosities   in    the    range   $\rm    \approx
10^{24}-10^{26}\,W/Hz$.   Therefore,  a  contribution  by  synchrotron
emission in  the optical/near-IR  wavebands is a  possibility.  Figure
\ref{fig_radiosed}  explores   this  scenario  by   extrapolating  the
observed  1.4\,GHz  flux  density   to  the  UV  assuming  a  standard
synchrotron    spectrum   of   the    form   $f_\nu\propto\nu^{-0.75}$
\citep[e.g.][]{Condon1992}.  This figure shows that the observed radio
emission  is too faint  to explain  the observed  red colours  of most
radio  detected  2MASS  QSOs,   in  agreement  with  previous  studies
\citep[e.g.][]   {Glikman2007}.   A   possible  exception   is  source
2MASS\_11.    This    lensed   source   was    first   identified   by
\cite{Gregg2002} because of its unusually red $B-K$ colour.  The radio
flux density of this system  is sufficiently bright ($\rm \approx 1 \,
Jy$) that  can potentially affect the optical  colours as  suggested by
\cite{Whiting2001}.

\cite{Richards2003} showed the observed  optical colours of QSOs are a
strong function of  redshift and proposed the use  of relative colours
(e.g.   $\rm \Delta (g-i)$;  difference between  the observed  and the
median colour  of QSOs at  the same redshift) to  discriminate between
intrinsically red  and dusty systems.   These authors showed  that the
$\rm \Delta (g-i)$ distribution of SDSS optically selected QSOs can be
approximated by  a Gaussian, suggesting  a range of  intrinsic optical
continuum slopes, and a red  tail, which is dominated by dust reddened
systems.   All the  2MASS QSOs  presented in  this paper  have $\Delta
(g-i)>0.8$, which  places them in the  red tail of  the SDSS optically
selected   QSO    distribution   \citep[e.g.    see    Figure   6   of
][]{Richards2003}  and in  the  region of  the  parameter space  where
sources with  dust reddening $E(B-V)\ga0.1$ are expected  to be found.
\cite{Richards2003}  shows  that  such  red colours  are  most  likely
because of dust rather than a steep optical power-law spectrum.

Dust reddening  indeed provides good fits to  the UV--to--near-IR SEDs
of  the   2MASS  QSOs,  if   the  SMC  extinction  curve   is  adopted
\citep{Richards2003}.  This  type of extinction  is strongly supported
by the shapes of the observed UV/optical SEDs, which are smooth and do
not show  the characteristic bump at rest-frame  $\rm \approx 2200\AA$
of  the  Galactic extinction  curve  \citep{Gordon2003}.  The  derived
optical extinctions for the 2MASS  QSOs are in the range $A_V=1.3-3.2$
(see Table  \ref{tab_restframe}), much higher  than optically selected
QSO samples,  which are typically sensitive to  systems with $A_V<0.8$
or      $\rm     E(B-V)<0.2$      for     SMC      type     extinction
\citep[e.g.][]{Richards2003}.

An independent  estimate of  the reddening can  be obtained  using the
Balmer or  Paschen line  decrements. For the  former we  assume Case-B
recombination  \citep{Storey1995},   while  for  the   latter  we  use
empirically determined ratios,  as Case B recombination is  not a good
approximation for the Paschen lines of QSOs \citep[e.g][]{Glikman2007,
Soifer2004}. This  excercise is  possible for 4  out of 10  2MASS QSOs
(see  Appendix).   These  values  typically  do  not  agree  with  the
reddening derived  from SED fitting.   A similar trend using  a larger
sample has been reported by \cite{Glikman2007}.  This implies that the
dust distribution is  complex and therefore the line  emission and the
continuum are absorbed by different material.

The ratio of the total  infrared luminosity in the dust torus ($3-1000
\mu m$)  to the reddening corrected  optical ($\rm 0.1 -3  \mu m$) QSO
luminosity, $L_{tor}/L_{opt}$,  has been  interpreted as a  measure of
the dust  covering factor  in AGN (Rowan-Robinson  et al.   2008). For
optically selected  QSOs in the  SWIRE survey the distribution  of the
$\log(L_{tor}/L_{opt}$) is well fitted  by a Gaussian with mean --0.10
and standard deviation 0.26 (Rowan-Robinson et al. 2008). We find that
6/8  2MASS  QSOs  with  spectrocopic  redshifts (see  Table  3),  have
$\log(L_{tor}/L_{opt})  > 0.0$,  suggesting covering  fractions larger
than the typical  SWIRE optically selected QSO. In  this comparison we
have excluded  the two 2MASS  QSOs without spectroscopic  redshifts to
avoid   uncertainties  in  the   determination  of   their  rest-frame
properties because of errors in the photometric redshift estimates.

In addition  to the AGN  activity, a number  of sources in  the sample
also show  evidence for star-formation.  There are  objects (see Table
\ref{tab_restframe})  with  excess emission  in  the  far-IR over  the
extrapolation  of  the  dust  torus  models  of  \cite{rowan1995}  and
\cite{Efstathiou1995}.   In our  modeling  of the  observed SEDs  this
excess is fit with cool  dust associated with starbursts. The total IR
luminosity  of  this  component   is  large,  in  the  range  $\approx
10^{12}-10^{13}\, L_\odot$ suggesting  high star-formation rates, $\rm
>  100 \, M_\odot\,yr^{-1}$.   Additionally, there  are sources  in the
sample  with excess  emission in  the  bluer SDSS  optical bands,  $g$
and/or  $u$,  over  the  expectation  of  the  reddened  QSO  template
(2MASS\_3,  2MASS\_10,  2MASS\_12).  This  excess  can  be  fit by  an
additional starburst  component that  dominates over the  reddened QSO
emission in the  UV part of the spectrum.   We cannot exclude however,
the  possibility of  scattered radiation  from the  AGN itself  as the
origin of this excess.  It is interesting nevertheless, that 2MASS\_3,
one of the objects that shows the UV excess, has narrow emission lines
in addition to the broad wings  of the $\rm H\alpha$.  The flux ratios
of the narrow emission line components place this source in the region
of Transition Objects in  the diagnostic diagram of \cite{Kewley2001},
suggesting some level of star-formation.

In  summary, the  red 2MASS  sources studied  in this  paper  are dust
reddened   QSOs,    while   a   large   fraction    (7/10   in   Table
\ref{tab_restframe})   have    far-IR   properties   consistent   with
star-formation in the host galaxy.

\section{Comparison with optically selected QSOs}

\subsection{Star-formation rate}

In this  section we first  explore possible systematic  differences in
the IR SEDs  of the red 2MASS QSOs with  optically selected ones.  For
this   comparison   we  use   the   Palomar-Green   (PG)  QSO   sample
\cite{Schmidt1983},   for  which   a  rich   set   of  multiwavelength
observations is  available.  We consider in  particular the sub-sample
of 64 PG QSOs with mid-  and far-IR observations ($\rm 5-200\mu m$) in
the ISO archive analysed  and presented by \cite{Haas2003}.  From that
sub-sample only sources with SDSS optical photometry (total of 26) are
used  here.  The near-IR  data for  these sources  are from  the 2MASS
survey.   In   order  to  avoid  uncertainties   associated  with  the
photometric redshift estimates, in this section we consider 2MASS QSOs
with spectroscopic redshifts  only. The median redshift of  the PG QSO
subsample used here is $\approx0.161$,  lower than that of 2MASS QSOs,
$z\approx0.5$.   

\begin{figure}
\begin{center}
 \rotatebox{0}{\includegraphics[height=0.9\columnwidth]{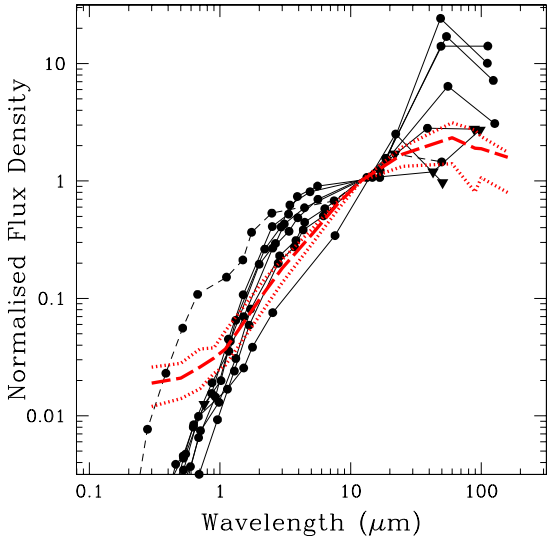}}
\end{center}
\caption{The rest-frame  SEDs of 2MASS  QSOs are shown with  the black
  dots connected with lines.  Only sources with spectroscopic redshift
  are  plotted.   The  long-dashed   thick  red  line  is  the  median
  rest-frame  SED of optically  selected QSOs.   The red  dotted lines
  correspond  to the 25th  and 75th  percentiles of  the distribution.
  Survival statistics as implemented  in the ASURV package  (Isobe et
  al.  1986;  Lavalley et al.  1992)  are used to  estimate the median
  and the 25th/75th percentiles of  the PG QSO SEDs taking into account
  upper limits in the flux density measurements of individual sources.
  All  SEDs  are normalized  to  the  rest-frame  $\rm 12\mu  m$  flux
  density, which provides a good  proxy to total AGN power independent
  of  dust  extinction. 2MASS  QSOs  have  much  steeper SEDs  in  the
  UV/optical, as  expected from dust reddening.  Also  some 2MASS QSOs
  have higher  far-IR flux density  (normalised to $\rm 12\mu  m$ flux
  density)  compared to  the median  SED of  PG QSOs.   The  2MASS QSO
  2MASS\_11 has significantly different SED from the rest of the 2MASS
  QSOs  in  the sample  and  is  marked  with the  short-dashed  black
  line.   This   is   a   lensed   radio   source   at   $z\approx2.2$
  \citep{Gregg2002}.  Synchrotron  emission may affect  the rest-frame
  UV/optical        SED       of        this        radio       source
  \citep[e.g.][]{Whiting2001}.}\label{fig_sed_all} 
\end{figure}

\begin{figure}
\begin{center}
 \rotatebox{0}{\includegraphics[height=0.9\columnwidth]{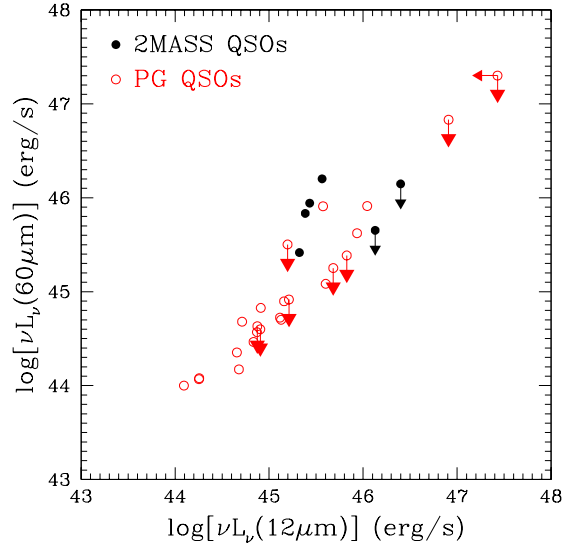}}
\end{center}
\caption{$\rm  60 \mu  m$  luminosity plotted  against  $\rm 12\mu  m$
  luminosity.   Only  2MASS   QSOs  with  spectroscopic  redshift  are
  shown. The rest-frame $\rm 60 \mu  m$ and $\rm 12\mu m$ flux density
  is estimated from the observed SED by linearly interpolating between
  data  points.    Two  out  of  eight  2MASS   sources  with  optical
  spectroscopy (2MASS\_08 and 2MASS\_11)  are at $z\approx2.2$ and the
  rest-frame  $\rm 60\mu  m$  is redshifted  outside the  Spitzer/MIPS
  photometric bands. An extrapolation would be uncertain and we choose
  not to estimate $\nu L_\nu  (60\mu m)$ for these sources.  This plot
  is similar to  that presented by Netzer et al.  (2007, their Fig. 5)
  and  Lutz  et  al. (2008,  their  Fig.   5)  and suggests  that  the
  star-formation  rate of  QSOs  (measured  by their  $\rm  60 \mu  m$
  luminosity) scales with  AGN power, measured here by  the $\rm 12\mu
  m$ luminosity.   There is tentative  evidence that 4/6 of  the 2MASS
  QSOs plotted in  the figure have more star-formation  for their AGN
  power    compared    to    optically    selected   PG    QSOs,    on
  average.}\label{fig_ll}
\end{figure}

Figure \ref{fig_sed_all} compares the  rest-frame UV-to-IR SEDs of the
2MASS QSOs with those of the optically selected PG QSOs.  The SEDs are
normalised  to the  rest-frame $\rm  12\mu  m$ flux  density, as  this
wavelength is proposed  to be nearly unbiased census  of the AGN power
\citep[e.g.][]{Spinoglio1989}.  As  expected dust reddening  makes the
UV/optical SEDs  of 2MASS QSOs steeper  than PG QSOs.   Also, there is
evidence that  the far-IR properties  of the two samples  differ.  The
normalised rest-frame far-IR flux density of at least 4 out of 8 2MASS
sources {\it with} spectroscopic redshifts is elevated compared to the
median SED  of PG  QSOs.   Our  SED  modeling  (see  Table
\ref{tab_restframe})  and recent  studies of  optically  selected QSOs
\citep[e.g][]{Haas2003,  Schweitzer2006},   suggest  that  the  far-IR
emission of these monsters is dominated by star-formation.  Under this
assumption,  Figure \ref{fig_sed_all}  therefore shows  that  at least
some 2MASS QSOs have, on average, higher star-formation than optically
selected QSOs.   This is  further demonstrated in  Figure \ref{fig_ll}
which plots  $\rm 60\,  \mu m$  luminosity against $\rm  12 \,  \mu m$
luminosity (see figure caption for details on the determination of the
luminosities). For given $\rm 12 \, \mu m$ luminosity, some 2MASS QSOs
are  more  luminous  at $\rm  60\,  \mu  m$.   The assessment  of  the
significance  of   this  excess  is   limited  by  the   small  sample
size. Nevertheless,  we use survival statistics as  implemented in the
ASURV  package \citep{Isobe1986,  Lavalley1992} to  estimate  the null
hypothesis probability  that the distribution of  the luminosity ratio
$\log \rm \nu L_\nu  (12\mu m) / \nu L_\nu (60\mu m)$  of 2MASS and PG
QSOs  is drawn  from the  same parent  population.  Using  the Gehan's
generalised Wilcoxon  test we  estimate this probability  to be  3 per
cent. The null hypothesis can therefore be rejected at the 97 per cent
level ($\rm \approx 2 \sigma$ in the case of Gauss distribution).

\cite{Schweitzer2006}  and \cite{Netzer2007}  find that  a substantial
fraction   of   optically  selected   PG   QSOs   show  evidence   for
star-formation (up  to 70\% based  on the far-IR detection  rate, 30\%
based on PAH detection rate).  Moreover, they found that indicators of
star-formation rate (e.g. PAH or $\rm 60\mu m$ luminosities) correlate
well with indicators  of AGN power (e.g.  $\rm  6\mu m$ or 5100\,\AA\,
luminosities) for these systems.  The red 2MASS QSOs studied here also
show evidence  for ongoing star-formation:  6 out of 8  sources (75\%)
with  spectroscopic  redshifts  are  fit by  an  additional  starburst
component in  the IR  (see Table \ref{tab_restframe}).   An additional
interesting  result from  our  analysis is  that  at the  97 per  cent
significance level  the 2MASS QSOs have excess  star-formation for their
AGN power compared to the optically selected PG QSOs. 

\subsection{Radio detection rate}

An interesting trend in Table 1 is the high fraction of the 2MASS QSOs
with detection at 1.4\,GHz: 6/10  are radio sources in either the NVSS
of the the FIRST surveys.  This is unexpected because the sources were
selected  independent  of  their  radio  properties,  unlike  previous
efforts to  identify reddened QSOs, which started  from radio selected
samples  to identify  objects  with very  red optical/near-IR  colours
\citep[e.g.][]{Gregg2002,    White2003,    Glikman2004,   Glikman2007,
  Urrutia2008a}.

Next  we explore whether  the radio  detection rate  of 2MASS  QSOs is
different from optically selected ones.  For this exercise, we compare
against the SDSS  QSOs to take advantage of the  large sample size and
the well defined selection criteria. We use in particular the SDSS-DR3
quasar   catalog  of  \cite{Schneider2005},   which  was   created  by
inspecting all spectra that were either targeted as quasar candidates,
or classified as  a quasar by the SDSS  spectroscopic pipelines.  This
catalogue includes objects from all categories of spectroscopic target
selection in the SDSS, not  just those selected as QSO candidates. For
our   purposes    we   use   only   targeted   SDSS    QSOs   in   the
\cite{Schneider2005} catalogue,  i.e. sources  that belong to  the the
main (or low-redshift) and the  high-redshift QSO samples of the SDSS.
These  samples include  QSOs  with $15<i<19$\,mag  that were  selected
based  on  the optical  colour  criteria  of \cite{Richards2002}.   We
further apply a cut $K<14.5$\,mag to match the magnitude limit used to
select   2MASS   QSOs.    There   are   638  such   sources   in   the
\cite{Schneider2005} catalogue.  The  $K=14.5$\,mag magnitude cut does
not introduce incompleteness in the sample, as it corresponds to $i \la
17.5$\,mag for the average  $(i-K)$ colour of optically selected QSOs,
which  takes  values  in   the  range  2--3,  depending  on  redshift.
Similarly,  the  bright  magnitude  limit $i>15$\,mag  for  follow  up
spectroscopy  of  the  SDSS  QSO  candidates  corresponds  to  $K  \ga
12.5$\,mag.  QSOs brighter  than  that  may be  missed  from the  SDSS
sample.  The surface density of such bright QSOs is likely to be small
however, and it  is not expected to introduce  any significant bias in
the SDSS optically selected QSO sample.

In order to estimate the radio detection rate of the sample we use the
FIRST radio  survey. There are  582 SDSS QSOs with  $K<14.5$\,mag that
fulfill  the \cite{Richards2002}  criteria and  overlap with  the area
covered  by  the  FIRST  survey.    A  total  of  172  of  them,  i.e.
$\approx30$  per   cent,  are  detected  at  1.4\,GHz   to  the  FIRST
limit. This  fraction should  be compared with  the detection  rate of
2MASS QSOs, i.e. 60 per  cent. Using binomial statistics we estimate a
4 per  cent probability of  at least 6  radio detections in  10 trials
given the  success rate of 30  per cent of the  optically selected QSO
sample.  The radio detection rate of 2MASS QSOs is higher than that of
optically selected  ones, albeit  at the 96  per cent  level ($\approx
2\sigma$).  This  is tentative evidence  that red 2MASS QSOs  are more
often radio  sources than optically  selected QSOs. \cite{Glikman2007}
have identified reddened 2MASS  QSOs by preselecting candidates in the
FIRST radio survey.  Their radio selected reddened sample represents a
substantial fraction of the  overall radio detected QSO population, up
to 60 per  cent. This finding further supports  an association between
red QSOs and radio emission. 

\begin{figure}
\begin{center}
 \rotatebox{0}{\includegraphics[height=0.9\columnwidth]{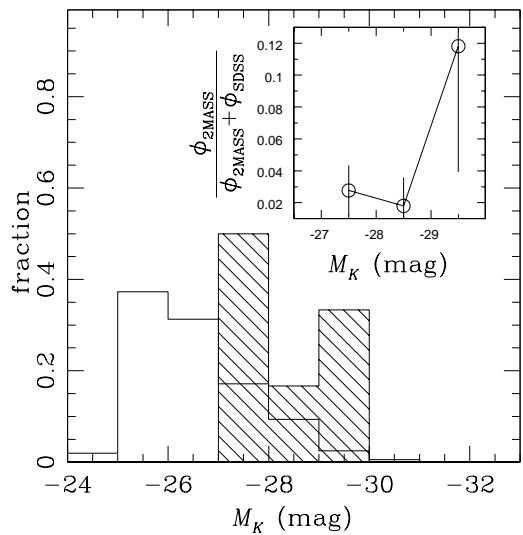}}
\end{center}
\caption{$M_K$  distribution of  optically selected SDSS QSOs with
$K<14.5$\,mag  (open  histogram)  and  reddened  2MASS  QSOs  (hatched
histogram). For the latter, $M_K$ is corrected for extinction. For both
samples we plot only sources  with redshift $z<1$. This way we exclude
high  redshift lensed  sources, such  as 2MASS\_11,  with artificially
enhanced  luminosities.  The  inset plot  shows as a  function of  $M_K$ the
fraction between the 2MASS and the total (2MASS + SDSS) QSO luminosity
functions  estimated using the  1/Vmax formalism  as described  in the
text. At bright luminosities reddened  2MASS QSO represent 12\% of the
overall population.}\label{fig_mkdist}
\end{figure}

We  caution that  2MASS QSOs  are more  luminous, on  average,  in the
$K$-band  than  the  SDSS   ones.   This  is  demonstrated  in  Figure
\ref{fig_mkdist} plotting  the $M_K$ distribution of  the two samples.
Although  the radio detection  rate of  SDSS QSOs  does not  appear to
depend on $M_K$ it might be possible that the somewhat different radio
detection rates of the two samples  are because of 2MASS QSOs are more
luminous.   Larger samples  of  reddened QSOs  are  needed to  further
explore  possible  differences in  their  radio  emission compared  to
optically selected samples.

\subsection{Fraction of reddened QSOs}

The fraction of  reddened QSOs within the overall  population is still
under debate with different studies  suggesting values from 15 to over
50   per  cent   \citep[e.g][]{Richards2002,   Wilkes2002,  White2003,
Glikman2004}.  The  difficulty in assessing this fraction  is that one
has to  take into account the  effect on dust in  the detectability of
reddened QSOs, which  in effect means that at  a given magnitude limit
one detects only the most  luminous reddened QSOs.  In this section we
take this effect into account  by estimating the maximum volume, Vmax,
that a source could be detected taking into account the dust reddening
and its luminosity.

First however,  we directly compare  the surface density of  the 2MASS
QSOs with those of optically selected ones. In the previous section we
counted  638 QSOs  in the  SDSS-DR3 catalogue  of \cite{Schneider2005}
with  $K<14.5$\,mag  that  fulfill the  \cite{Richards2002}  criteria.
Using  the spectroscopic  area of  SDSS-DR3, $\rm  3732 \,  deg^2$, we
estimate a sky density  of $\rm 0.17\,deg^{-2}$ for optically selected
QSOs to $K=14.5$\,mag.  In contrast, we find 10 reddened 2MASS QSOs in
the $\rm  5282\, deg^2$ of  the photometric SDSS-DR3  catalogue, which
corresponds  to a  surface  density of  $\rm 0.002\,deg^{-2}$,  2\,dex
lower than optically selected  QSOs.  The reddened 2MASS QSOs selected
here appear to  be very rare and therefore  insignificant, in terms of
numbers, for  the overall  population of QSOs.   However, as  shown in
Figure \ref{fig_mkdist}  reddened QSOS are more luminous  and the SDSS
QSOs and it is therefore not surprising they are rarer.

We    address    this     issue    using    the    1/Vmax    formalism
\citep[e.g.][]{Lilly1995} to estimate the $K$-band luminosity function
of the 2MASS QSOs relative to that of the overall QSO population.  For
each source with intrinsic absolute magnitude $M_K$ the maximum volume
Vmax is  estimated given  the selection function  of the  sample, i.e.
$K<14.5$\,mag,   $R-K>4$  and   $J-K>2$.   For   the   calculation  of
magnitudes, colours and k-corrections  we adopt the SDSS composite QSO
SED   \citep{VandenBerk2001}  and   redden   it  by   SMC  type   dust
\citep{Pei1992}  assuming  the optical  extinction,  $A_V$, listed  in
Table \ref{tab_restframe}.  We  then sum the 1/Vmax of  each source in
$M_K$  luminosity  bins.   The  same  procedure is  repeated  for  the
optically selected SDSS QSOs  with $K<14.5$\,mag, assuming $A_V=0$ for
this population. In  order to minimise evolution effects  and to avoid
high  redshift  sources  for  which lensing  artificially  boosts  the
observed luminosity (e.g. 2MASS\_11),  we only consider 2MASS and SDSS
QSOs with  $z<1$. The results  are shown in  the inset plot  of Figure
\ref{fig_mkdist}, which plots as a  function of $M_K$ the ratio of the
2MASS sources  luminosity function over the  total luminosity function
of  both  2MASS and  SDSS  QSOs.  The  fraction  of  reddened QSOs  is
increasing  from $\approx  3\pm2$  per cent  at  $M_K=-27.5$ to  about
$12\pm8$ per cent  of the overall population at  $M_K=-29.5$.  This is
evidence that  reddened QSOs become important  at bright luminosities.
We  caution  that  small  number  statistics  are  an  issue  in  this
calculation.  Larger  reddened QSO samples at  magnitudes fainter than
the 2MASS $K=14.5$ limit are  needed to further explore this trend and
to   provide   a  more   reliable   measure   of   the  reddened   QSO
fraction. Efforts to identify QSOs  in the the UKIRT Infrared Deep Sky
Survey (UKIDSS, $K\approx18$\,mag)  including dust reddened ones, have
recently appeared  in the literature  \citep{Maddox2008}, although the
source selection is very different to that presented here.
 
\section{Discussion}

This  paper combines  photometric data  from the  largest  optical and
near-IR   surveys,  SDSS   and  2MASS,   to  select   QSOs   with  red
optical/near-IR  colours,   which  are   atypical  to  those   of  the
traditional population  of optically selected broad-line  QSOs.  It is
shown that  the UV/optical SEDs  of the red  QSOs can be  explained by
moderate amounts of dust  extinction, $A_V=1.3-3.2$.  An open question
is whether  red QSOs  can be explained  by orientation  arguments, and
therefore are a subset of optically selected QSOs, or they represent a
different stage of the life of QSOs.

On the  latter point,  there is  a large volume  of literature  on the
possible evolution link between  starbursts, QSOs and spheroids.  Both
observations  \citep[e.g.][]{Sanders1988,  Sanders1996,  Clements2000,
Canalizo2001,      Komossa2003,      Alexander2005,      Veilleux2006,
Schweitzer2006, Dasyra2007} and numerical simulations of major mergers
\citep[e.g.][]{DiMatteo2005,  Springel2005, Hopkins2006}  support such
an association.   It is proposed  in particular, that  some mechanism,
possibly mergers, drive gas efficiently to the central galaxy regions,
thereby triggering powerful starbursts and black hole growth at a high
accretion  rate.   During  this  stage  both the  central  engine  the
star-forming  regions are  enshrouded in  dust and  gas clouds.   As a
result the  system appears  luminous in the  infrared (LIRGs  or ULIRG
stage).  AGN and/or starburst  driven winds (feedback) will eventually
develop and blow away the  dust cocoon at the nuclear regions, thereby
allowing  the AGN to  shine unobscured.  This is  the stage  where the
system  can be  identified as  UV/optically luminous  broad  line QSO.
When  all the  cold gas  in the  nuclear regions  is  depleted, either
because it is  used up in star-formation, consumed  by the black-hole,
heated or pushed away by the feedback processes, the AGN and starburst
will switch off,  leaving behind a quiescent galaxy.   In this picture
reddened QSOs  are systems  captured just before  or during  the final
blow-out stage of their evolution.

Our findings, although  limited by the small number  of sources in the
sample,  are  in  broad  agreement  with this  scenario.   The  far-IR
properties of  about 70 per cent  of the reddened  QSOs are consistent
with   starburst  activity.    There   is  also   evidence  that   the
star-formation rate of these systems,  normalised to the AGN power, is
higher than that  of optically selected QSOs, on  average.  A possible
interpretation  of this  finding  is that  feedback  from the  central
engine has not yet suppressed the star-formation in the host galaxy at
the same level  as in typical QSOs. There  is also tentative evidence,
significant at  the 96 per  cent confidence level, that  reddened QSOs
are  more often associated  with radio  sources compared  to optically
selected QSOs.  If this is confirmed by larger samples, it may suggest
the onset of the AGN driven  feedback in these systems via radio jets.
The   tentative   evidence   for   larger  dust   covering   fractions
(approximated  by $L_{tor}/L_{opt}$)  of the  2MASS QSOs  presented in
this  paper  compared  to  optically  selected  SWIRE  QSOs,  is  also
consistent with the young AGN interpretation.  With the data presented
here we cannot comment on  the optical morphology of reddened QSO host
galaxies.  \cite{Hutchings2003,  Hutchings2006} and \cite{Urrutia2008}
explored  this  issue using  a  different  sample  of red  2MASS  QSOs
selected to  have near-IR colours $J-K>2$\,mag.  They  report a higher
incidence of  systems with interaction signs in  their sample compared
to  optically selected  QSOs  \citep[but see][]{Marble2003}.  Although
this  conclusions  is  limited   by  the  absence  of  an  appropriate
comparison  sample,   it  is   consistent  with  the   merger  induced
starburst/AGN activity picture above,  and suggests that reddened QSOs
are  at an earlier  evolutionary stage  than optically  selected ones.
The relatively low  fraction of 2MASS QSOs, i.e. 3-12  per cent of the
overall QSO  population depending on $M_K$, may  therefore reflect the
short timescale  of the particular evolutionary stage  relative to the
QSO lifetime.

Another  class  of sources  that  are  also  proposed to  precede  the
optically  luminous QSO  phase are  ULIRGs with  broad  emission lines
\citep[type-I  ULIRGs;   e.g.][]{Kawakatu2003,  Kawakatu2006}.   These
systems have young stellar  populations \citep[$\la \rm 300 \,Myr$ old
][]{Canalizo2001},   a  large  fraction   of  them   experience  tidal
interactions \citep{Canalizo2001, Lipari2005}, their black hole masses
are smaller  compared to ellipticals  and optically selected  QSOs for
the size  of their bulge \citep{Kawakatu2006},  Broad Absorption Lines
(BAL)  indicative   of  outflows   are  frequent  in   this  population
\citep{Canalizo2001}, and most of  them have narrow permitted emission
  lines  (e.g.  FWHM  of $\rm  H\beta$ less  than $\rm  4000 \,  km \,
  s^{-1}$) similar  to narrow-line Seyfert 1 galaxies,  which are also
  believed to be young  AGN \citep{Zheng2002, Anabuki2004}.  The 2MASS
  QSOs studied  here share  some of the  properties of  type-I ULIRGs,
  i.e.  $L_{IR} >10^{12}\,L\odot$,  permitted emission lines with FWHM
  typically  $\rm  <2000\,km\,s^{-1}$,  ongoing star-formation.  Also,
  \cite{Urrutia2008a} have  recently found an  unusually high fraction
  of BAL sources  among red 2MASS QSOs selected  with criteria similar
  to those used here.  It is  therefore tempting to link the two types
  of sources.  A major difference between the two populations however,
  is that Type-I ULIRGs are typically less reddened at UV/optical than
  2MASS QSOs  \citep[e.g.][]{Canalizo2001}.  Is it  therefore possible
  that 2MASS QSOs are even younger than type-I ULIRGs in the evolution
  scheme outlined above?

\begin{figure}
\begin{center}
 \rotatebox{0}{\includegraphics[height=0.9\columnwidth]{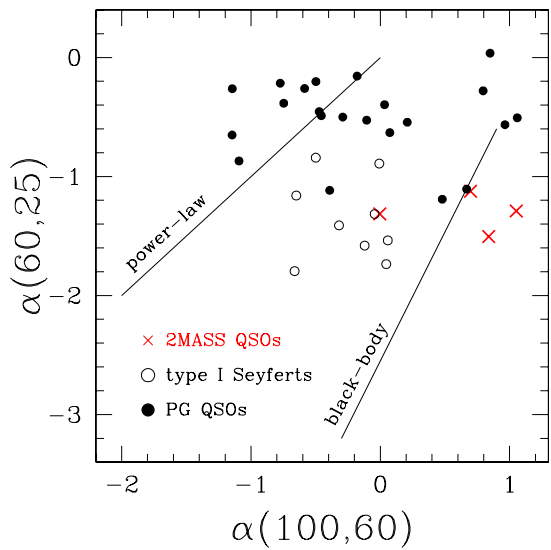}}
\end{center}
\caption{Infrared colour-colour plot.   The colour indices are defined
$\alpha(\lambda_1,\lambda_2) =  - \log (f_{\lambda_1}/f_{\lambda_1}) /
\log (\lambda_1/\lambda_2)$,  where $f_{\lambda}$ is  the flux density
at wavelength $\lambda$. The filled  circles are the sample of PG QSOs
used here, open circles are  the type-I ULIRGS of Canalizo \& Stockton
(2001) and the (red) crosses are the low redshift 2MASS QSOs ($z<0.6$)
in the present sample (total of 4).  For higher redshift 2MASS QSOs at
least one of  the rest-frame wavelengths 25, 60 or  $\rm 100\mu m$, is
outside the  Spitzer spectral window.  For  the 2MASS and  PG QSOs the
rest-frame flux densities used to estimate the infrared colour indices
are estimated by interpolating between the data points of the observed 
SEDs. The infrared  colours of the Canalizo \&  Stockton (2001) sample
are estimated using  the IRAS flux densities. The  diagonal lines show
the   position  of  power-law   and  black-body   SEDs  \citep[adapted
from][]{Lipari1994}.  }\label{fig_ir_colour}
\end{figure}

Figure \ref{fig_ir_colour}  explore this possibility  by comparing the
infrared  colours,  $\alpha(60,25)$  against  $\alpha(100,  60)$  (see
figure caption  for details on the  estimation of the  colours), of PG
and  2MASS  QSOs with  the  sample of  type-I  ULIRGs  of Canalizo  \&
Stockton (2001).   Different types  of sources occupy  fairly distinct
regions in  this figure. Sources dominated  by cool dust  are found in
the  left  part   of  the  diagram,  close  to   line  marked  ``black
body''. Systems  dominated by  non-thermal emission populate  the left
part of  the diagram, close  to the ``power-law'' line.   According to
the evolution  scenario described above QSOs should  evolve from right
to  left in  Figure \ref{fig_ir_colour}.   Optically selected  PG QSOs
dominate  the upper left  part of  the diagram,  the type-I  ULIRGS of
Canalizo  \& Stockton  (2001)  are  selected to  lie  between the  two
diagonal  lines, while  the  four low  redshift  2MASS QSOs  ($z<0.6$)
scatter toward the  far right end of the  diagram.  This suggests, but
does not  prove, that  2MASS QSOs may  represent young  QSOs, possibly
before they  become type-I ULIRGs and  eventually typical UV/optically
luminous broad  line QSOs.  Clearly, more data  are needed  to further
explore this scenario.  What fraction of 2MASS QSOs qualify for narrow
line  Seyfert 1s? What  are the  black hole  masses and  the accretion
rates  of these  systems?  Do  we  see optical  spectral evidence  for
outflows  (e.g.  absorption  features), similar  to those  observed in
type-I ULIRGs  \citep[e.g.][]{Canalizo2001}?  What are  the properties
of  their host  galaxies?  All these  are  questions that  need to  be
addressed before concluding on the nature of 2MASS QSOs.

\section{Conclusions}

We combine optical and near-IR photometry from the ground with Spitzer
mid- and  far-IR broadband data to explore  the SEDs of a  sample of 10
QSOs with  red optical and  near-IR colours selected by  combining the
2MASS with the SDSS. The main conclusions are

\begin{itemize}

\item The UV/optical part of the  SED can be fit assuming an intrinsic
spectrum similar  to that of typical optically  selected QSOs reddened
by moderate amounts of dust.

\item  There is  evidence that  red 2MASS  QSOs have  higher  level of
star-formation  (approximated by  the  $\rm 60\mu  m$ luminosity)  for
their AGN power  (measured by the $\rm 12\mu  m$ luminosity), compared
to optically selected PG QSOs.

\item We  find that  2MASS QSOs are  more often associated  with radio
emission,  compared to  optically  selected SDSS  QSOs.  This  excess,
although interesting, is only significant at the 96\% level

\item The  fraction of  red QSOs in  the overall  population increases
  from    $3\pm2$\%    at    $M_K=-27.5$\,mag   to    $12\pm8$\%    at
  $M_K=-29.5$\,mag,   suggesting  that   these  systems   become  more
  important at the bright luminosities.

\item We argue that these findings are consistent with a picture where
  reddened QSOs are young AGN shortly before they shed their cocoon of
  dust  and  gas  clouds  and  become  typical  UV/optically  luminous
  QSOs.  Comparison with  Seyfert-1 ULIRGs,  another class  of sources
  proposed  to be  young QSOs,  suggests  that red  2MASS QSOs  likely
  represent an earlier evolutionary stage.

\end{itemize}

\section{Acknowledgements}

We thank the anonymous referee for providing constructive comments and
suggestions that significantly improved this paper. This work has been
supported   by   funding  from   the   Marie-Curie  Fellowship   grant
MEIF-CT-2005-025108  (AG), the  STFC  (DLC) and  the National  Science
Foundation through grant AST-0507781 (MSB). This publication makes use
of data products from the Two  Micron All Sky Survey, which is a joint
project of the University of Massachusetts and the Infrared Processing
and Analysis Center/California Institute  of Technology, funded by the
National Aeronautics and Space Administration and the National Science
Foundation. The  United Kingdom Infrared Telescope is  operated by the
Joint  Astronomy  Centre  on  behalf  of the  Science  and  Technology
Facilities Council of  the U.K. The WHT and  its service programme are
operated on  the island of La Palma  by the Isaac Newton  Group in the
Spanish Observatorio  del Roque de  los Muchachos of the  Instituto de
Astrofísica de Canarias.

\appendix

\section{Notes on Individual Sources}

{\bf 2MASS QSO\#2:}  The optical spectrum of this  sources shows broad
Balmer emission lines, $\rm H\beta$, $\rm H\gamma$. Fitting a Gaussian
to the $\rm  H\beta$ we estimate a FWHM of  $\rm \approx2000\, km\, s^
{-1}$. The [NeIII]\,3869 line is  also seen in the spectrum with broad
profile. It is  interesting that the Balmer lines  are redshifted from
the   [OIII]\,4959   and  5007\AA\,   lines   by  $\rm   \approx1500\,
km\,s^{-1}$.  A similar effect was reported by McIntosh et al.  (1999)
for luminous  high redshift  QSOs.  In their  sample the  mean redward
offset of the $\rm H\ beta$  line with respect to the [OIII]\,5007 was
estimated to be  $\rm \, \approx 520 km  \, s^{-1}$. Visual inspection
of  the optical  spectrum also  suggests  that the  Balmer lines  have
asymmetric profiles with blue wings.

\noindent {\bf  2MASS QSO\#3:} This source has  narrow emission lines,
$\rm H\beta$, [OIII]\,4959 and  5007\AA\, [NII]\,6583\AA, and the $\rm
H\alpha$ line with broad wings.  A simultaneous two component Gaussian
fit to the  blended $\rm H\alpha$ and [NII]\,6583\AA\,  lines fails to
reproduce  the flux  in the  wings  of $\rm  H\alpha$.  An  additional
gaussian is  introduced to  account for the  broad wings. The  FHWM of
this component is estimated  to be $\rm \approx 1500\,km\,s^{-1}$.  We
also estimate diagnostic emission  line ratios using the $\rm H\beta$,
[OIII]\,5007\AA\, [NII]\,6583\AA, and the narrow component of the $\rm
H\alpha$.   We estimate $\rm  \log ([OIII]  / H\beta)=+0.50$  and $\rm
\log  ([NII]  / H\alpha)=  -0.58$,  which  place  this source  in  the
Transition   Objects    region   of   the    diagnostic   diagram   of
\cite{Kewley2001}. The  radio emission from this  source is consistent
with  the  starburst  radio/far-IR correlation  of  \cite{Condon1992}.
Adopting the SED of M\,82  to extrapolate the measured rest-frame $\rm
60\,\mu m$  flux density to $  \rm 100\, \mu m$  we estimate $q=2.19$,
which  is  consistent with  the  median  for starbursts  $q=2.3\pm0.2$
\citep{Helou1985,  Condon1992}.   We estimate  an  extinction of  $\rm
A_V=1.6$  from  the  Balmer  decrement  of the  narrow  emission  line
components. This  compares well to  $A_V=1.9$ determined from  the SED
fits to the continuum.
   
\noindent {\bf 2MASS QSO\#4:}  Only narrow emission lines are observed
in the optical spectrum of this source, $\rm H\beta$, [OIII]\,4959 and
5007\AA. The spectral  window does not cover the  $\rm H\alpha$ or the
Paschen  line region  which suffer  less extinction.   This  source is
therefore likely to also have broad lines.

\noindent {\bf  2MASS QSO\#5:}  This source has  optical spectroscopic
observations, which  however did  not yield a  redshift determination.
SED fitting estimates a photometric redshift $z=3.1$.

\noindent  {\bf  2MASS QSO\#6:}  We  find  broad  $\rm H\alpha$,  $\rm
Pa\alpha$ and $\rm Pa\beta$ emission lines. A three component Gaussian
fit  to  the $\rm  H\alpha$  line (to  account  for  the blended  $\rm
[NII]\,6583\,\AA$ line) gives a FWHM  of $\rm 2200\, km \, s^{-1}$ for
the  broad component.  The  FWHM of  the Paschen  lines is  about $\rm
1500\, km  \, s^{-1}$, estimated by  single Gaussian fits  to the line
profiles.    The  extinction   is  $\rm   A_V=5.7$  from   the  Balmer
decrement. In  this calculation we  use the narrow line  $\rm H\alpha$
component  to  compare  with   the  narrow  $\rm  H\beta$  line.   The
extinction above is higher than $A_V=2.3$ determined from the SED fits
to  the   continuum.   The  Paschen  line  ratio   is  estimated  $\rm
Pa\alpha/Pa\beta=1.9$  and is consistent  with $A_V=5.3$  or $A_V=9.6$
adopting  respectively, an empirically  determined intrinsic  ratio of
1.05 \citep{Soifer2004} or 0.64 \citep{Glikman2006}.

\noindent {\bf  2MASS QSO\#7:} Optical  spectroscopic observations are
available, but  did not yield  a redshift determination.   SED fitting
estimates a photometric redshift $z=2.73$.

\noindent {\bf  2MASS QSO\#8:}  This source lies  at $z=2.16$  and has
broad  $\rm H\alpha$ with  FWHM of  about $\rm  4700\, km  \, s^{-1}$,
estimated by single Gaussian fit  to the line profile.  The extinction
is $\rm A_V=4.4$ from the Balmer decrement. In this calculation we use
single guassian line fits to the $\rm H\alpha$ and $\rm H\beta$ lines.
The extinction above is higher  than the $A_V=1.2$ determined from the
SED fits to the continuum.

\noindent  {\bf 2MASS QSO\#10:}  The optical  spectrum of  this source
shows  narrow  emission  lines  only,  [OII]\,3727\AA,  $\rm  H\beta$,
[OIII]\,4959  and 5007\AA\. The  near-IR spectrum  of this  source has
been presented by Glikman et  al. (2004) and shows broad $\rm Pa\beta$
emission line.

\noindent {\bf 2MASS  QSO\#11:} This is the lensed  reddened QSO first
found  by Gregg  et al.   (2002).   The optical  and near-IR  spectrum
presented in that paper shows broad emission lines.

\noindent {\bf  2MASS QSO\#12:} This  source has broad  $\rm Pa\alpha$
and $\rm Pa\beta$ emission lines with  FWHM of about $\rm 1600\, km \,
s^{-1}$, estimated by  single Gaussian fits to the  line profiles. The
Paschen  line  ratio  is  estimated  $\rm  Pa\alpha/Pa\beta=1.7$.  The
corresponding reddening is $A_V=4.3$ or $A_V=8.6$ for the empirically
determined  intrinsic  ratios  of  1.05  \citep{Soifer2004}  and  0.64
\citep{Glikman2006}  respectively. These  ratios are  larger  than the
reddening of $A_V=3$ determined from the SED fits to the
continuum. This source has  also been selected in the sample of
Glikman et  al. (2007) and Urrutia et al. (2008a). 

\bibliography{mybib}{}
\bibliographystyle{mn2e}

\end{document}